\documentclass[11pt,a4paper]{article}
\usepackage{jcappub,natbib}
\usepackage{braket}
\input{colordvi.tex}

\title{Optimal limits on primordial magnetic fields from CMB temperature bispectrum of passive modes}
\author[a]{Maresuke Shiraishi,}
\author[b]{Daisuke Nitta,}
\author[a]{Shuichiro Yokoyama}
\author[a]{and Kiyotomo Ichiki}
\affiliation[a]{Department of Physics and Astrophysics, Nagoya University, 
Nagoya 464-8602, Japan} 
\affiliation[b]{Astronomical Institute, Tohoku University, Aoba Aramaki Aoba, Sendai 980-8578, Japan}
\emailAdd{mare@nagoya-u.jp}
\emailAdd{nitta@astr.tohoku.ac.jp}
\emailAdd{yokoyama.shuuichirou@h.nagoya-u.jp}
\emailAdd{ichiki@a.phys.nagoya-u.ac.jp}
\abstract{
We investigate bounds on the strength of the primordial magnetic field (PMF) from the cosmic microwave background (CMB) bispectra of the intensity (temperature) modes induced from the auto- and cross-correlated bispectra of the scalar and tensor components of the PMF anisotropic stress. At first, we construct a general formula for the CMB intensity and polarization bispectra from PMFs composed of any type of perturbation. Then we derive an approximate expression which traces the exact shape of the CMB bispectrum in order to reduce the computation time with respect to a large number of the multipole configurations, and also show that the non-Gaussian structure coming from PMFs is classified as the local-type configuration. 
Computing the signal-to-noise ratio on the basis of the approximate formula with the information of the instrumental noises and resolutions, 
we find expected upper bounds on the magnetic field strength, when the magnetic spectrum is nearly scale invariant ($n_B = -2.9$), 
smoothed on $1 {\rm Mpc}$ scale at 95\% confidence level from the WMAP and PLANCK experiments as $B_{1 \rm Mpc} < 4.0 - 6.7 {\rm nG}$ and $3.8 - 6.5 {\rm nG}$, respectively, depending on the energy scale of the magnetic field production from $10^{14} {\rm GeV}$ to $10^3 {\rm GeV}$. Our new consequences imply slight overestimations by the previous rough discussions.}
\arxivnumber{1201.0376}
\begin{document}
\maketitle
\flushbottom
\allowdisplaybreaks[4]


\def\up{\;\raise1.0pt\hbox{$'$}\hskip-6pt\partial\;}
\def\down{\;\overline{\raise1.0pt\hbox{$'$}\hskip-6pt\partial\;}}
\section{Introduction}

Thanks to the precise measurements of the cosmological and astrophysical events, there exist some inevitable evidences of the existence of 
the cosmological-scale magnetic fields, whose strength is about a few microgauss, in galaxies and clusters of galaxies at redshift $z \sim 0.7 - 2.0$ \cite{Bernet:2008qp, Wolfe:2008nk, Kronberg:2007dy}. In addition, recently, from the data of the Fermi and HESS experiments, it has been found that the magnetic field strength in the inter-galactic medium is larger than ${\cal O}(10^{-15}) G$ \cite{Neronov:1900zz, Dolag:2010ni, Tavecchio:2010mk}. In order to realize the current magnetized Universe via the astrophysical magnification known for the dynamo mechanism (e.g., refs.~\cite{Grasso:2000wj, Widrow:2002ud, Giovannini:2003yn}), a lot of theories predict the existence of the seed magnetic field in the early Universe (e.g., refs.~\cite{Grasso:2000wj, Bamba:2006ga, Martin:2007ue, Demozzi:2009fu}). This seed field acts as a source of the cosmic microwave background (CMB) and generates characteristic patterns on its anisotropy. Therefore, the attempts to extract the properties of the seed field from the CMB two-point correlation function (power spectrum) have been well studied, and we had a broad understanding that the Universe favors the scale-invariant magnetic field whose magnitude (normalized by the present value) is less than ${\cal O}(1) {\rm nG}$ (see e.g., refs.~\cite{Durrer:1998ya, Mack:2001gc, Lewis:2004ef, Yamazaki:2008gr, Paoletti:2008ck, Finelli:2008xh, Shaw:2009nf, Yamazaki:2010nf, Shaw:2010ea, Paoletti:2010rx}).

On the other hand, even if the primordial magnetic field (PMF) obeys the Gaussianity, due to the quadratic dependence of the CMB fluctuation on them, the higher-order correlation of the CMB fluctuations also appear \cite{Brown:2005kr}. In particular, since the CMB three-point function (bispectrum) has been well utilized as a measure of the primordial non-Gaussianity (e.g., refs.~\cite{Bartolo:2004if, Komatsu:2010fb, Komatsu:2010hc}), there are a lot of literature which have been analyzing the impacts of the PMF on the CMB bispectrum and seeking its reasonable constraints \cite{Seshadri:2009sy, Caprini:2009vk, Trivedi:2010gi, Cai:2010uw, Shiraishi:2010yk, Shiraishi:2011fi, Shiraishi:2011dh}. However, in refs.~\cite{Seshadri:2009sy, Caprini:2009vk, Trivedi:2010gi, Cai:2010uw}, the authors approximately computed the integrals over the wave number vectors in the CMB bispectrum of the magnetic scalar modes. In our works \cite{Shiraishi:2010yk, Shiraishi:2011fi, Shiraishi:2011dh}, despite success of the exact calculation of the CMB auto-correlated bispectra between the vector and tensor modes, bounds on the PMF magnitude were estimated only by a rough comparison with the CMB bispectrum from curvature perturbations. In this sense, these estimations may have involved several kind of uncertainties. In addition, the previous studies have not focused on the contribution of the cross-correlated bispectra between the different perturbation modes yet, that is, the previous works seem to be incomplete. 

The goal of this paper is to obtain bounds on the PMF strength by computing of the signal-to-noise ratio including all information of the scalar and tensor passive modes if the PMF spectrum has the nearly scale-invariant shape as $n_B = -2.9$. Here, the passive modes denote the CMB signals induced by the growth of the metric perturbations due to the PMF anisotropic stress prior to neutrino decoupling, and dominate over the CMB fluctuation from PMFs for $\ell \lesssim 1000$ \cite{Lewis:2004ef, Shaw:2009nf}. To do this, we firstly give an exact formula for the CMB intensity and polarization bispectrum composed of any type of perturbation. However, as the numerical computation based on this formula requires a great deal of time, we present a faster form for the auto- and cross-correlated bispectrum between the scalar and tensor modes, which traces the shape of the exact one, by an appropriate approximation similar to the way of ref.~\cite{Lyth:2006gd}. Then, we also find that the bispectrum of the PMF anisotropic stresses represents the local-type non-Gaussianity. After that, using this approximate formula, we compute the signal-to-noise ratio in terms of the intensity-intensity-intensity bispectra with the effects of the instrumental noises and resolutions of the current observations as WMAP and PLANCK \cite{Komatsu:2010fb, :2006uk}, and obtain new bounds on the PMF strength. 

This paper is organized as follows. In the next section, we review the definition and statistical properties of the PMF. In section~\ref{sec:CMB_bis}, we produce exact and approximate formulae for the CMB bispectrum, and analyze these behaviors. In section~\ref{sec:SN}, we present the signal-to-noise ratio and the limits on the PMF strength. The final section is devoted to a summary and discussion of this paper. 

Throughout this paper, we obey the definition of the Fourier transformation as 
\begin{eqnarray}
f({\bf x}) \equiv \int \frac{d^3 {\bf k}}{(2 \pi)^3} \tilde{f}({\bf k}) e^{i {\bf k} \cdot {\bf x}}~,
\end{eqnarray}
and the rule that the subscripts and superscripts of the Greek characters and alphabets run from 0 to 3 and from 1 to 3, respectively.

\section{Statistics for the primordial magnetic fields}

In this section, according to ref.~\cite{Shiraishi:2011fi}, let us summarize the statistical properties of the stochastic primordial magnetic fields (PMFs), ${\bf B}({\bf x}, \tau)$, on the homogeneous background Universe characterized by the Friedmann-Lema{\^\i}tre-Robertson-Walker metric as 
$ds^2 = a(\tau)^2 ( - d\tau^2 + \delta_{bc} dx^b dx^c )$,
where $\tau$ is a conformal time and $a (\tau)$ is a scale factor. The
expansion of the Universe makes the amplitude of the magnetic fields
decays as $1 / a^2$ and hence we can draw off the time dependence as
${\bf B}({\bf x}, \tau) = {\bf B}({\bf x}) / a^2$. Using this scaling relation, each component of the energy momentum tensor of the PMF is given by 
\begin{eqnarray}
\begin{split}
T^0_{~0}(x^\mu) &= - \rho_B = - \frac{1}{8 \pi a^4} B^2({\bf x}) \equiv - \rho_\gamma(\tau) \Delta_B(x^\mu)~, \\
T^0_{~c}(x^\mu) &= T^b_{~0}(x^\mu) = 0~, \\
T^b_{~c}(x^\mu) &= \frac{1}{4\pi a^4} \left[\frac{B^2({\bf x})}{2}\delta^b_{~c}
 -  B^b({\bf x})B_c({\bf x})\right] 
\equiv \rho_\gamma(\tau) 
\left[ \Delta_B(x^\mu) \delta^b_{~c} + \Pi^b_{Bc}(x^\mu) \right]~, \label{eq:EMT_PMF} 
\end{split}
\end{eqnarray}
where we have introduced the photon energy density $\rho_{\gamma}$ 
to confine the time dependence of $a^{-4}$. 
After the Fourier transformation, the spatial parts are described as
\begin{eqnarray}
\begin{split}
T^b_{~c}({\bf k},\tau) &\equiv {\rho_\gamma}(\tau)
\left[ \delta^b_{~c} \Delta_B({\bf k}) + \Pi_{Bc}^b({\bf k}) \right]~, \\
\Delta_B({\bf k}) &= {1 \over 8\pi \rho_{\gamma,0}}
\int \frac{d^3 {\bf k}'}{(2\pi)^3} 
B^b({\bf k'}) B_b({\bf k} - {\bf k'}) ~, \\
\Pi^b_{Bc}({\bf k}) &=-{1 \over 4\pi \rho_{\gamma,0}} \int \frac{d^3
 {\bf k'}}{(2 \pi)^3} B^b({\bf k'}) B_c({\bf k} - {\bf k'})~, 
\label{eq:EMT_PMF_fourier} 
\end{split}
\end{eqnarray}
 where $\rho_{\gamma, 0}$ denotes the present energy density of photons. In the following
discussion, for simplicity of calculation, we ignore the trivial
time-dependence. Hence, the index can be lowered by 
$\delta_{bc}$ and the summation is implied for repeated indices.

Assuming that ${\bf B}({\bf x})$ obeys the Gaussianity, its power spectrum is defined by 
\footnote{Here we neglect the helical term.}
\begin{eqnarray}
\langle B_a({\bf k})B_b({\bf p})\rangle
= (2\pi)^3 {P_B(k) \over 2} P_{ab}(\hat{\bf k}) \delta ({\bf k} + {\bf p})~. \label{eq:def_pmf_power}  
\end{eqnarray}  
Here, we use the projection tensor, which comes from the divergence free nature of PMFs, as 
\begin{eqnarray}
P_{ab}(\hat{\bf k}) 
\equiv \sum_{\sigma = \pm 1} \epsilon^{(\sigma)}_a(\hat{\bf k}) \epsilon^{(-\sigma)}_b(\hat{\bf k}) 
 =  \delta_{ab} - \hat{k}_a \hat{k}_b~, \label{eq:projection}
\end{eqnarray}
with $\hat{\bf k}$ and $\epsilon_a^{(\pm 1)}$ being a unit vector and a normalized divergenceless polarization vector which satisfies the orthogonal
condition; $\hat{k}^a \epsilon_a^{(\pm 1)} = 0$. The details of the
relations and conventions of the polarization vector are described in
appendix~\ref{appen:polarization}.  Although the form of the power spectrum $P_B(k)$ strongly depends on the production mechanism of the PMF, we assume a simple power law shape given by 
\begin{eqnarray}
P_B(k) = A_B k^{n_B}~,
\end{eqnarray}
where $A_B$ and $n_B$ denote the amplitude and spectral index of the
power spectrum of PMFs, respectively.  In order to
parametrize the strength of the PMF, we smooth the magnetic fields with a
conventional Gaussian filter on a comoving scale $r$:
\begin{eqnarray}
B_r^2 \equiv \int_0^\infty \frac{k^2 dk}{2 \pi^2} e^{-k^2 r^2} P_B(k) ~.
\end{eqnarray}
Then, $A_B$ is calculated as
\begin{eqnarray}
{A}_B = {
\left(2 \pi \right)^{n_B + 5} B_r^2 \over \Gamma(\frac{n_B + 3}{2}) k_r^{n_B + 3} }~,
\end{eqnarray}
where $\Gamma(x)$ is the Gamma function and $k_r \equiv 2 \pi / r$.

As we will mention in the next section, the anisotropic stress of the PMF, $\Pi_{B ab}$, acts as a source of the CMB fluctuation and hence we shall focus on its statistical properties. Using a projection tensor on the scalar mode $O_{ab}^{(0)} \equiv - \hat{k}_a \hat{k}_b + \delta_{ab}/3$, $\epsilon_a^{(\pm 1)}$, and a transverse-traceless polarization tensor $e_{ab}^{(\pm 2)} \equiv \sqrt{2} \epsilon_a^{(\pm 1)} \epsilon_b^{(\pm 1)} $ (for more detail see appendix \ref{appen:polarization}), the fluctuation of the PMF anisotropic stress is decomposed into the scalar, vector and tensor parts:
\begin{eqnarray}
\begin{split}
\Pi_{Bs}^{(0)}({\bf k}) 
&= \frac{3}{2} O_{ab}^{(0)}(\hat{\bf k}) \Pi_{B ab}({\bf k})  ~,\\ 
\Pi_{Bv}^{(\pm 1)}({\bf k}) 
&= \frac{1}{2}
\left[ \hat{k}_a {\epsilon}_b^{(\mp 1)}(\hat{\bf k})  + \hat{k}_b {\epsilon}_a^{(\mp 1)}(\hat{\bf k}) \right] \Pi_{B ab}({\bf k}) 
= \hat{k}_a {\epsilon}_b^{(\mp 1)}(\hat{\bf k}) \Pi_{B ab}({\bf k})  ~, \\
\Pi_{Bt}^{(\pm 2)}({\bf k}) 
&= \frac{1}{2} e_{ab}^{(\mp 2)}(\hat{\bf k}) \Pi_{B ab}({\bf k})  ~, 
\label{eq:PMF_ani}
\end{split}
\end{eqnarray}
where in the second equality of the second equation, we use a fact that $\Pi_{B ab}$ is symmetric for the permutation of $a$ and $b$. 
According to equation~(\ref{eq:EMT_PMF}), the PMF anisotropic stress at an arbitrary point, $\Pi^{a}_{Bb} ({\bf x})$, depends quadratically on the Gaussian PMF at that point, ${\bf B}({\bf x})$. This situation is identical to the local-type non-Gaussianity of curvature perturbations: $\zeta({\bf x}) \equiv \zeta_{\rm G}({\bf x}) + (3/5) f_{\rm NL}^{\rm local} 
\left[ \zeta_{\rm G}^2({\bf x}) - \Braket{\zeta_{\rm G}^2({\bf x})} \right]$ with $\zeta, \zeta_{\rm G}$ and $f_{\rm NL}^{\rm local}$ being the curvature perturbation, its Gaussian part and the nonlinear parameter of the local-type non-Gaussianity, respectively \cite{Komatsu:2001rj}. Hence, it is expected that the statistical properties of the PMF is similar to those of the local-type non-Gaussianity. This will be discussed in the next section. 

Using equation~(\ref{eq:def_pmf_power}) and the Wick's theorem, the bispectrum of the PMF anisotropic stresses is calculated as
\begin{eqnarray}
\Braket{\Pi_{Bab} ({\bf k_1}) \Pi_{Bcd} ({\bf k_2})
 \Pi_{B ef} ({\bf k_3})} 
&=&\left( - 4\pi \rho_{\gamma,0} \right)^{-3} 
\left[ \prod_{n=1}^3 
\int_0^{k_D} k_n'^2 dk_n' P_B(k_n') \int d^2 \hat{\bf k_n'} \right]
\nonumber \\
&& \times 
\delta({\bf k_1} - {\bf k_1'} + {\bf k_3'}) 
\delta({\bf k_2} - {\bf k_2'} + {\bf k_1'}) 
\delta({\bf k_3} - {\bf k_3'} + {\bf k_2'}) \nonumber \\
&& \times \frac{1}{8} [P_{ad}(\hat{\bf k_1'}) P_{be}(\hat{\bf k_3'})
P_{cf}(\hat{\bf k_2'}) + \{a \leftrightarrow b \ {\rm or} \ c \leftrightarrow d \ {\rm or} \ e \leftrightarrow f\}], \nonumber \\ 
\label{eq:bis_EMT}
\end{eqnarray}
where $k_D$ is the Alfv\'en-wave damping length scale
\cite{Jedamzik:1996wp, Subramanian:1997gi} as $k_D^{-1} \sim {\cal
O}(0.1)\rm Mpc$ and the curly bracket denotes the symmetric seven terms
under the permutations of indices: $a \leftrightarrow b$, $c
\leftrightarrow d$, or $e \leftrightarrow f$. Here, for convenience of the calculation, we express as a more symmetric form than that of ref.~\cite{Brown:2005kr}. To avoid the divergence of
$\Braket{\Pi_{Bab} ({\bf k_1}) \Pi_{Bcd} ({\bf k_2}) \Pi_{Bef} ({\bf
k_3})}$ in the IR limit, the spectral index is
limited as $n_B > -3$. Note that equation~(\ref{eq:bis_EMT}) is the six-point correlation of the Gaussian variables, that is, the higher-order correlation than the bispectrum of curvature perturbations. Hence, for formulating the CMB bispectrum induced by PMFs, appropriate treatments of the complicated computation corresponding to the one-loop diagram are required. 

\section{CMB scalar, vector and tensor bispectrum} \label{sec:CMB_bis}

In this section, we present the exact and approximate forms for the CMB bispectrum of the intensity and polarization modes induced by the scalar, vector and tensor perturbations of the non-Gaussian PMF anisotropic stresses, and discuss the consistency between the exact and approximate bispectra via numerical computations. 

Conventionally, the intensity, and $E$- and $B$-mode polarizations ($X \equiv I, E, B$) generated from the CMB scalar, vector and tensor anisotropies ($Z \equiv S, V, T$) are expanded by the spin-$0$ spherical harmonics as 
\begin{eqnarray}
\frac{\Delta X^{(Z)}(\hat{\bf n})}{X} = \sum_{\ell m} a_{X, \ell m}^{(Z)} Y_{\ell m}(\hat{\bf n})~,
\end{eqnarray}
where $\hat{\bf n}$ is a line-of-sight direction. Then, the each-mode coefficient, $a_{X, \ell m}^{(Z)}$, is formally given by 
\begin{eqnarray}
a^{(Z)}_{X, \ell m} = 4\pi (-i)^\ell 
\int \frac{d^3{\bf k}}{(2\pi)^3} 
\sum_{\lambda} [{\rm sgn}(\lambda)]^{\lambda+x} \xi^{(\lambda)}({\bf k}) 
{\cal T}_{X, \ell}^{(Z)}(k) {}_{-\lambda}Y_{\ell m}^*(\hat{\bf k})  ~, 
\label{eq:alm_PMF_general}
\end{eqnarray}
where $\lambda$ expresses the helicities of the scalar, vector and tensor perturbations as $\lambda = 0$ for $Z = S$, $\lambda = \pm 1$ for $Z = V$ and $\lambda = \pm 2$ for $Z = T$, $x$ discriminates the parities of the intensity and two polarizations as $x = 0$ for $X = I, E$ and $x = 1$ for $X = B$, and $\xi^{(\lambda)}$ and ${\cal T}_{X, \ell}^{(Z)}(k)$ are the primordial perturbation and transfer function of each mode, respectively.\footnote{Here, we set $0^0 = 1$.} 
If PMF exists, its anisotropic stress acts as the source of the CMB anisotropy. For example, the primordial gravitational waves and curvature perturbations are amplified by the PMF anisotropic stresses prior to neutrino decoupling although these growths stop due to the compensation by the neutrino anisotropic stresses after neutrino decoupling. These extra gravitational waves and curvature perturbations survive passively, and generate additional fluctuations on the tensor- and scalar-mode CMB fields, respectively, similar to the non-magnetic ones on large and intermediate scales \cite{Lewis:2004ef, Kojima:2009gw, Shaw:2009nf}.\footnote{In refs.~\cite{Bonvin:2011dt, Bonvin:2011dr}, by taking into account the effects in both the inflationary and the radiation-dominated epochs consistently, the authors concluded that the solution of the curvature perturbation spoils. See however ref.~\cite{Barnaby:2012tk}.} On the other hand, in the vector mode, as the gravitational potential decays away via the Einstein equation, additional CMB fluctuation arises from not the above mechanism but the enhancement of the vorticity of photons by the Lorentz force from the PMF at small scales \cite{Durrer:1998ya, Mack:2001gc, Lewis:2004ef}. This Lorentz force is also originated from the PMF anisotropic stress, hence, we can express each-mode primordial perturbations with the PMF anisotropic stress: 
\begin{eqnarray}
\begin{split}
\xi^{(0)}({\bf k}) &\approx 
- R_\gamma \ln\left(\frac{\tau_\nu}{\tau_B}\right) 
\Pi^{(0)}_{Bs}({\bf k}) 
 ~,\\
\xi^{(\pm 1)}({\bf k}) &\approx \Pi^{(\pm 1)}_{Bv}({\bf k})   ~,\\
\xi^{(\pm 2)}({\bf k}) &\approx 
6 R_\gamma \ln\left(\frac{\tau_\nu}{\tau_B}\right)  
\Pi^{(\pm 2)}_{Bt}({\bf k}) ~. \label{eq:initial_perturbation}
\end{split}
\end{eqnarray}
Here, note that $\xi^{(0)}$ and $\xi^{(\pm 2)}$ correspond to the superhorizon solutions of the Einstein equations with respect to the curvature perturbation and gravitational wave,\footnote{$\xi^{(\pm 2)}$ denotes each helicity state of the primordial gravitational wave, which is expressed as 
\begin{eqnarray}
\xi^{(\pm 2)}({\bf k}) = \frac{1}{2} e_{ij}^{(\mp 2)}(\hat{\bf k}) h_{ij}({\bf k})~, 
\end{eqnarray}
with $h_{ij} \equiv \delta g_{ij} / a^2$ being the spatial components of the metric perturbation.} respectively \cite{Lewis:2004ef, Kojima:2009gw, Shaw:2009nf, Barnaby:2012tk}, and depend on the production time of the PMF, $\tau_B$, the epoch of neutrino decoupling, $\tau_\nu \simeq 1 {\rm MeV}^{-1}$, and the ratio by the energy density of photons to all relativistic particles, $R_\gamma \sim 0.6$ 
for $\tau < \tau_\nu$. In this paper, as the upper and lower values of $\tau_B^{-1}$, we take the energy scale of the grand unification ($10^{14} {\rm GeV}$) to the electroweak symmetry breaking ($10^3 {\rm GeV}$), corresponding to $\tau_\nu / \tau_B = 10^{17}$ and $10^6$, respectively. 
As ${\cal T}_{X, \ell}^{(S)}$ and ${\cal T}_{X, \ell}^{(T)}$, we can use the standard cosmological transfer functions without the magnetic dependence \cite{Zaldarriaga:1996xe, Hu:1997hp, Weinberg:2008zzc, Shiraishi:2010sm} because posterior to neutrino decoupling, the PMF contributes little to the evolution of cosmological perturbations \cite{Shaw:2009nf}. In contrast, as ${\cal T}_{X, \ell}^{(V)}$, we should use the form involving the impacts of the PMF shown in Refs.~\cite{Durrer:1998ya, Mack:2001gc, Shiraishi:2011fi}. 

For convenience of the computation for the CMB bispectrum, we shall perform the multipole expansion of each-mode primordial perturbation as 
\begin{eqnarray}
\xi^{(\lambda)}({\bf k}) = \sum_{\ell m} \xi^{(\lambda)}_{\ell m}(k) {}_{-\lambda}Y_{\ell m}(\hat{\bf k})~. \label{eq:xi_expand}
\end{eqnarray}
Then, equation~(\ref{eq:alm_PMF_general}) changes to 
\begin{eqnarray}
a^{(Z)}_{X, \ell m} = 4\pi (-i)^\ell 
\int \frac{k^2 dk}{(2\pi)^3} 
\sum_{\lambda} [{\rm sgn}(\lambda)]^{\lambda+x} \xi^{(\lambda)}_{\ell m}(k) 
{\cal T}_{X, \ell}^{(Z)}(k)~, 
\end{eqnarray} 
and their bispectrum is easily formulated as
\begin{eqnarray}
\Braket{\prod^3_{n=1} a_{X_n, \ell_n m_n}^{(Z_n)}}
&=& \left[ \prod^3_{n=1}
4\pi (-i)^{\ell_n}
\int_0^\infty {k_n^2 dk_n \over (2\pi)^3}
\mathcal{T}^{(Z_n)}_{X_n, \ell_n}(k_n) \sum_{\lambda_n}
 [{\rm sgn}(\lambda_n)]^{\lambda_n + x_n} \right] \nonumber \\
&&\times \Braket{\prod_{n=1}^3 \xi_{\ell_n m_n}^{(\lambda_n)}(k_n)}~. 
\label{eq:CMB_bis_PMF_general} 
\end{eqnarray}
In the following subsections, by expanding $\Braket{\prod_{n=1}^3 \xi_{\ell_n m_n}^{(\lambda_n)}(k_n)}$ and performing summations over $\lambda_1, \lambda_2$, and $\lambda_3$, we firstly find an exact expression of the CMB bispectra from PMFs. However, it takes an awful lot of time to compute the CMB bispectra based on the exact form. Thus, using an approximate expression of the bispectrum of the PMF anisotropic stresses, we develop a faster form that can reconstruct the exact shape in section~\ref{subsec:pole_app}. 

\subsection{Exact expression} \label{subsec:exact}

\begin{figure}[t]
\begin{tabular}{cc}
\begin{minipage}[t]{0.5\hsize}
  \begin{center}
    \includegraphics[width=5cm,clip]{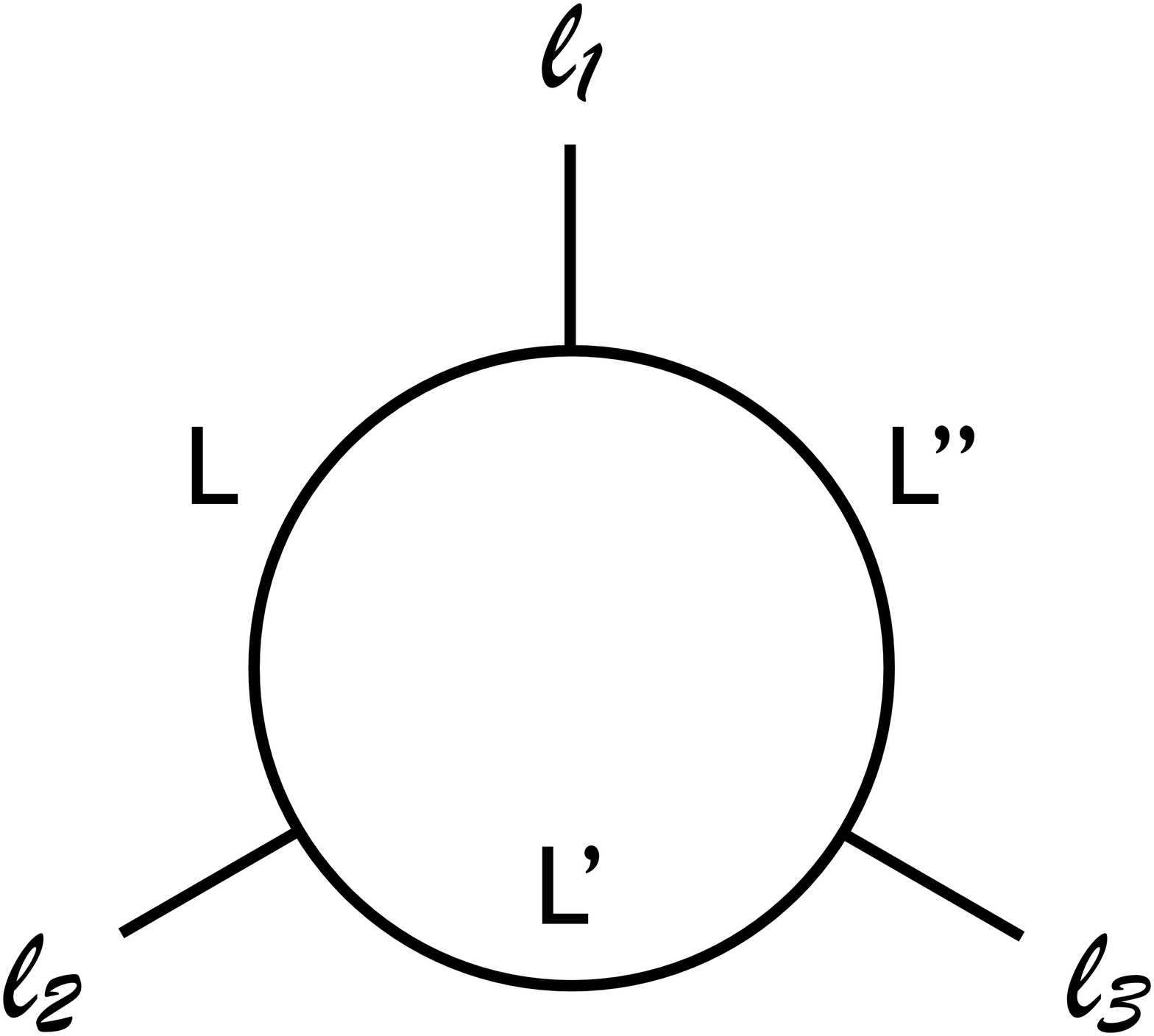}
  \end{center}
\end{minipage}
\begin{minipage}[t]{0.5\hsize}
  \begin{center}
    \includegraphics[width=5cm,clip]{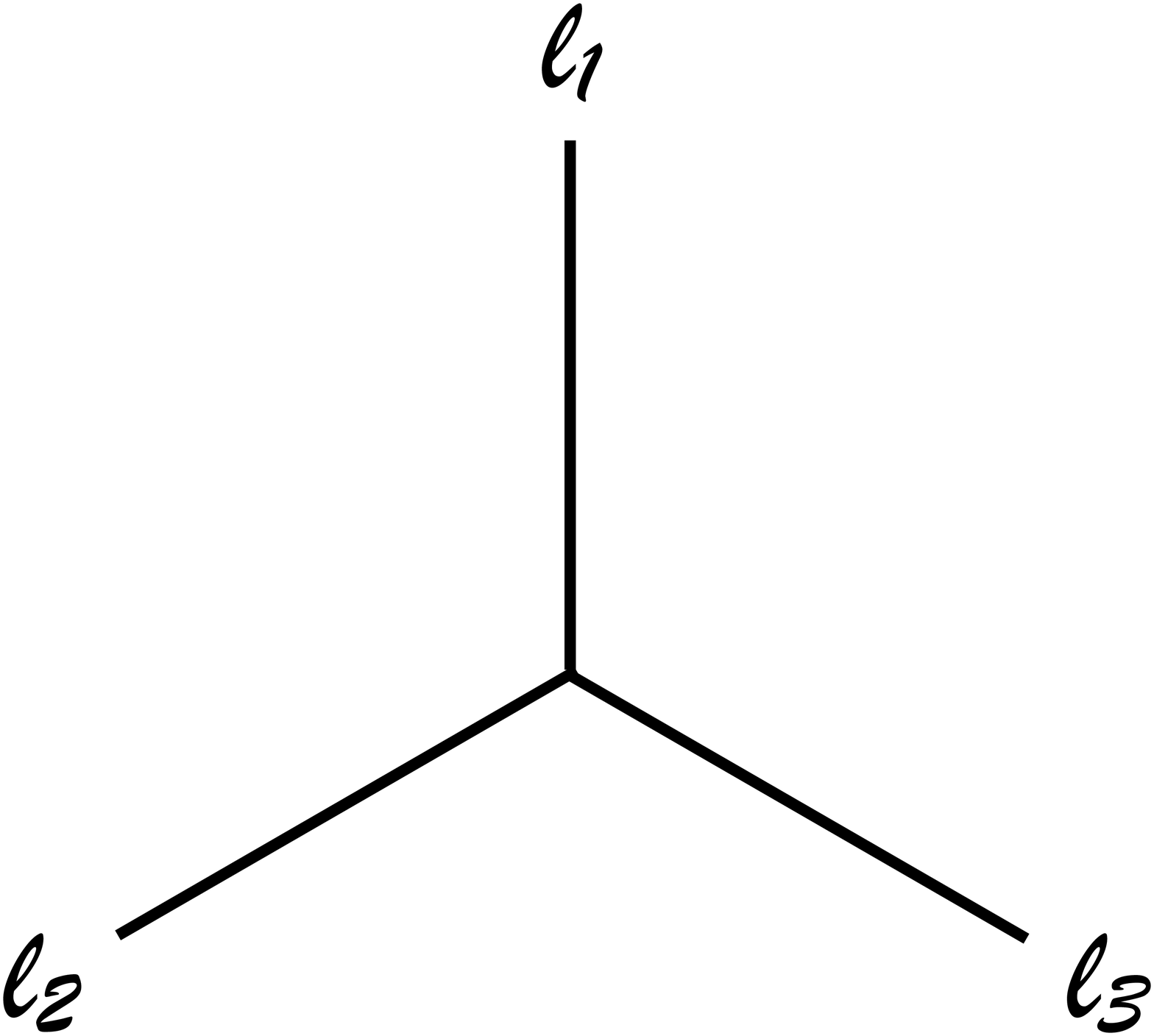}
  \end{center}
\end{minipage}
\end{tabular}
  \caption{Diagrams with respect to the multipoles. The left panel corresponds to the multipole configuration of equation~(\ref{eq:CMB_bis_PMF_complete}). Due to the Wigner-$6j$ symbol originated with the sextuplicate dependence on the Gaussian PMFs, $\ell_1, \ell_2$ and $\ell_3$ are linked with the dummy ones $L, L'$, and $L''$, and the one-loop structure is realized. The right panel represents the tree-structure diagram, which arises from the CMB bispectrum induced by the four-point function of the Gaussian seed fields.}
\label{fig:diagram}
\end{figure}

Here, we want to expand the CMB bispectrum (\ref{eq:CMB_bis_PMF_general}) with the mathematically rigid method \cite{Shiraishi:2011fi}. 
To obtain an analytic formula for the CMB bispectrum, we firstly should reduce the bispectrum of $\xi_{\ell m}^{(\lambda)}(k)$, which is explicitly given by 
\begin{eqnarray}
\Braket{\prod_{n=1}^3 \xi_{\ell_n m_n}^{(\lambda_n)}(k_n)}
=  \left[\prod_{n=1}^3 \int d^2\hat{\bf k_n} {}_{-\lambda_n}Y_{\ell_n m_n}^*(\hat{\bf k_n}) \right] 
\Braket{\prod_{n=1}^3 \xi^{(\lambda_n)}({\bf k_n})}~.
\end{eqnarray}
Note that $\Braket{\prod_{n=1}^3 \xi^{(\lambda_n)}({\bf k_n})}$ is composed of the bispectrum of the PMF anisotropic stresses (\ref{eq:bis_EMT}) and the relations (\ref{eq:PMF_ani}) and (\ref{eq:initial_perturbation}). 
To perform three angular integrals in this equation, we expand the delta function and projection tensors in $\Braket{\prod_{n=1}^3 \xi^{(\lambda_n)}({\bf k_n})}$ by using the spin-weighted spherical harmonics as
\begin{eqnarray}
\delta\left( \sum_{n=1}^3 {\bf k_n} \right) 
&=& 8 \int_0^\infty y^2 dy 
\left[ \prod_{n=1}^3 \sum_{L_n M_n} 
 (-1)^{L_n/2} j_{L_n}(k_n y) 
Y_{L_n M_n}^*(\hat{\bf k_n}) \right] \nonumber \\
&&\times 
I_{L_1 L_2 L_3}^{0~0~0}
\left(
  \begin{array}{ccc}
  L_1 & L_2 & L_3 \\
  M_1 & M_2 & M_3 
  \end{array}
 \right)~, \label{eq:delta}
\end{eqnarray}
and some equations described in appendix~\ref{appen:polarization}, respectively \cite{Shiraishi:2011fi, Shiraishi:2010kd}. 
Here, $j_l$ is the Bessel function and the $I$ symbol is defined by
\begin{eqnarray}
I^{s_1 s_2 s_3}_{l_1 l_2 l_3}
\equiv \sqrt{\frac{(2 l_1 + 1)(2 l_2 + 1)(2 l_3 + 1)}{4 \pi}}
\left(
  \begin{array}{ccc}
  l_1 & l_2 & l_3 \\
  s_1 & s_2 & s_3
  \end{array}
 \right)~.
\end{eqnarray}
Furthermore, we replace the angular integrals of the spin spherical
harmonics with the Wigner-$3j$ symbols, and sum up these
Wigner-$3j$ symbols over the azimuthal quantum numbers
in the same manner as in ref.~\cite{Shiraishi:2011fi}. After these complicated treatments, we can obtain a formula for the initial bispectrum composed of any combination of three kind of the primordial perturbations as 
\begin{eqnarray}
\Braket{\prod_{n=1}^3 \xi_{\ell_n m_n}^{(\lambda_n)}(k_n)}
&=& \left(
  \begin{array}{ccc}
  \ell_1 & \ell_2 & \ell_3 \\
  m_1 & m_2 & m_3
  \end{array}
 \right)
(-4 \pi \rho_{\gamma,0})^{-3}\left[ \prod_{n=1}^3 \int_0^{k_D} k_n'^2 dk_n' P_B(k_n') \right]
\nonumber \\ 
&& \times \sum_{L L' L''} \sum_{S, S', S'' = \pm 1} 
\left\{
  \begin{array}{ccc}
  \ell_1 & \ell_2 & \ell_3 \\
  L' & L'' & L 
  \end{array}
 \right\} \nonumber \\
&&\times 
f^{S'' S \lambda_1}_{L'' L \ell_1 }(k_3',k_1',k_1) f^{S S' \lambda_2}_{L L' \ell_2}(k_1',k_2',k_2)
f^{S' S'' \lambda_3}_{L' L'' \ell_3}(k_2',k_3',k_3)~,
 \label{eq:xi_bis_pmf_exact}
\end{eqnarray}
where 
\begin{eqnarray}
f^{S'' S \lambda}_{L'' L \ell}(r_3, r_2, r_1) 
&=& \sum_{L_1 L_2 L_3} \int_0^\infty y^2 dy j_{L_3}(r_3 y) j_{L_2}(r_2 y) j_{L_1}(r_1 y)  \nonumber \\
&& \times
 (-1)^{\ell + L_2 + L_3} (-1)^{\frac{L_1 + L_2 + L_3}{2}} 
I^{0~0~0}_{L_1 L_2 L_3} I^{0 S'' -S''}_{L_3 1 L''} I^{0 S -S}_{L_2 1 L} 
I_{L_1 \ell 2}^{0 \lambda -\lambda} 
 \left\{
  \begin{array}{ccc}
  L'' & L & \ell \\
  L_3 & L_2 & L_1 \\
  1 & 1 & 2
  \end{array}
 \right\} \nonumber \\
&&\times
\left\{
 \begin{array}{ll}
 - \frac{2}{\sqrt{3}} (8\pi)^{3/2} R_\gamma 
\ln \left(\tau_\nu / \tau_B\right)  & (\lambda = 0) \\
 \frac{2}{3} (8\pi)^{3/2} \lambda  & (\lambda = \pm 1) \\
  - 4 (8\pi)^{3/2} R_\gamma \ln \left( \tau_\nu / \tau_B \right) &
   (\lambda = \pm 2)
 \end{array}
\right. ~. \label{eq:f_exact}
\end{eqnarray}
Here, we stress that this expression involves all other modes than the vector-vector-vector and tensor-tensor-tensor ones which have been considered in our previous works \cite{Shiraishi:2010yk, Shiraishi:2011fi, Shiraishi:2011dh}. 
Note that the confinement of $m_1, m_2$ and $m_3$ to the Winger-$3j$ symbol 
$\left(
  \begin{array}{ccc}
  \ell_1 & \ell_2 & \ell_3 \\
  m_1 & m_2 & m_3
  \end{array}
 \right)$ ensures the rotational invariance of the CMB bispectrum, hence, in the following discussion, we focus on the angle-averaged form $B_{X_1 X_2 X_3, \ell_1 \ell_2 \ell_3}^{(Z_1 Z_2 Z_3)}$ defined by 
\begin{eqnarray}
\Braket{\prod_{n=1}^3 a_{\ell_n m_n}^{(Z_n)}}
\equiv 
\left(
  \begin{array}{ccc}
  \ell_1 & \ell_2 & \ell_3 \\
  m_1 & m_2 & m_3
  \end{array}
 \right) B_{X_1 X_2 X_3, \ell_1 \ell_2 \ell_3}^{(Z_1 Z_2 Z_3)}~.
 \label{eq:def_angle_averaged}
\end{eqnarray}
 Substituting equation~(\ref{eq:xi_bis_pmf_exact}) into equation~(\ref{eq:CMB_bis_PMF_general}), we can obtain an explicit form of the angle-averaged bispectrum as
\begin{eqnarray}
B_{X_1 X_2 X_3, \ell_1 \ell_2 \ell_3}^{(Z_1 Z_2 Z_3)}
&=& 
\left[ \prod\limits^3_{n=1}
(-i)^{\ell_n}
\int {k_n^2 dk_n \over 2\pi^2}
\mathcal{T}_{X_n, \ell_n}^{(Z_n)}(k_n)
\sum\limits_{\lambda_n} [{\rm sgn}(\lambda_n)]^{\lambda_n + x_n}
\int_0^{k_D} k_n'^2 dk_n' P_B(k_n') \right] \nonumber \\ 
&& \times 
(- 4 \pi \rho_{\gamma,0})^{-3} \sum_{L L' L''} \sum_{S, S', S'' = \pm 1} 
\left\{
  \begin{array}{ccc}
  \ell_1 & \ell_2 & \ell_3 \\
  L' & L'' & L 
  \end{array}
 \right\} \nonumber \\
&& \times
f^{S'' S \lambda_1}_{L'' L \ell_1}(k_3',k_1',k_1) f^{S S' \lambda_2}_{L L' \ell_2}(k_1',k_2',k_2)
f^{S' S'' \lambda_3}_{L' L'' \ell_3}(k_2',k_3',k_3)~. \label{eq:CMB_bis_PMF_complete}
\end{eqnarray}
From this equation, we can find that the CMB bispectrum from PMFs is not obtained unless performing three summations with respect to the additional multipoles $L, L'$, and $L''$, which is correlated with $\ell_1, \ell_2$ and $\ell_3$ via the Wigner-$6j$ symbol 
$\left\{
  \begin{array}{ccc}
  \ell_1 & \ell_2 & \ell_3 \\
  L' & L'' & L 
  \end{array}
 \right\}$. These extra summations come from the sextuplicate dependence on the Gaussian PMFs. This situation, which corresponds to the one-loop diagram as shown in the left panel of figure~\ref{fig:diagram}, leads to a lengthy computation.
In the next subsection, using an appropriate approximation, we reduce this loop calculation to the tree-level one (corresponding to the right diagram of figure~\ref{fig:diagram}). 

For the numerical calculation, we need to decompose the $f$ functions in equation~(\ref{eq:CMB_bis_PMF_complete}) and perform summations over $S, S', S'', \lambda_1, \lambda_2$ and $\lambda_3$. Using the analytic results as 
\begin{eqnarray}
\begin{split}
& \sum_{S, S', S'' = \pm 1} (-1)^{L_2 + L'_2 + L''_2 + L_3 + L'_3 + L''_3}
I^{0 S -S}_{L'_3 1 L} I^{0 S -S}_{L_2 1 L} I^{0 S' -S'}_{L''_3 1 L'} 
I^{0 S' -S'}_{L'_2 1 L'} I^{0 S'' -S''}_{L_3 1 L''} I^{0 S'' -S''}_{L''_2 1 L''}  \\
&\qquad\qquad = I_{L'_3 1 L}^{0 1 -1} I_{L_2 1 L}^{0 1 -1} I_{L''_3 1 L'}^{0 1 -1}
I_{L'_2 1 L'}^{0 1 -1} I_{L_3 1 L''}^{0 1 -1} I_{L''_2 1 L''}^{0 1 -1} \\
&\qquad\qquad\quad\times 
\left\{
 \begin{array}{ll}
 8 & (L'_3 + L_2, L''_3 + L'_2, L_3 + L''_2 = {\rm even}) \\
 0 & ({\rm otherwise}) 
 \end{array}
\right. ~, \\
& \sum_{\lambda_1 \lambda_2 \lambda_3} 
[{\rm sgn(\lambda_1)}]^{x_1} I^{0 \lambda_1 -\lambda_1}_{L_1 \ell_1 2} 
[{\rm sgn(\lambda_2)}]^{x_2} I^{0 \lambda_2 -\lambda_2}_{L'_1 \ell_2 2} 
[{\rm sgn(\lambda_3)}]^{x_3} I^{0 \lambda_3 -\lambda_3}_{L''_1 \ell_3 2} \\
&\qquad\qquad = I_{L_1~\ell_1~2}^{0 |\lambda_1| -|\lambda_1|}
 I_{L_1'~\ell_2~2}^{0 |\lambda_2| -|\lambda_2|} I_{L_1''~\ell_3~2}^{0
 |\lambda_3| -|\lambda_3|}  \\
&\qquad\qquad\quad\times 
\left\{
 \begin{array}{ll}
 2^{3-N_S} & (L_1 + \ell_1 + x_1, L_1' + \ell_2 + x_2,  L_1'' + \ell_3 +
  x_3 = {\rm even}) \\ 
 0 & ({\rm otherwise}) 
 \end{array}
\right. ~, \label{eq:summation_helicities}
\end{split}
\end{eqnarray} 
with $N_S$ being the number of the scalar modes constituting the CMB bispectrum,\footnote{Caution about a fact that $|\lambda|$ is determined by $Z$, namely, $|\lambda| = 0, 1, 2$ for $Z = S, V, T$, respectively.} equation (\ref{eq:CMB_bis_PMF_complete}) is rewritten as 
\begin{eqnarray}
B^{(Z_1 Z_2 Z_3)}_{X_1 X_2 X_3, \ell_1 \ell_2 \ell_3}
&=& 
C_{Z_1} C_{Z_2} C_{Z_3} 
\left( -4 \pi \rho_{\gamma,0} \right)^{-3} 
\sum _{L L' L''} 
\left\{
  \begin{array}{ccc}
  \ell_1 & \ell_2 & \ell_3 \\
  L' & L'' & L 
  \end{array}
 \right\}
\nonumber \\
&& \times
\sum_{\substack{L_1 L_2 L_3 \\ L'_1 L'_2 L'_3 \\ L''_1 L''_2 L''_3}} 
(-1)^{\sum_{n=1}^3\frac{L_n+L'_n+L''_n+2 \ell_n}{2}} 
I^{0~0~0}_{L_1 L_2 L_3} I^{0~0~0}_{L'_1 L'_2 L'_3} I^{0~0~0}_{L''_1
L''_2 L''_3}  \nonumber \\
&& \times
\left\{
  \begin{array}{ccc}
  L'' & L & \ell_1 \\
  L_3 & L_2 & L_1 \\
  1 & 1 & 2
  \end{array}
 \right\}
\left\{
  \begin{array}{ccc}
  L & L' & \ell_2 \\
  L'_3 & L'_2 & L'_1 \\
  1 & 1 & 2
  \end{array}
 \right\}
\left\{
  \begin{array}{ccc}
  L' & L'' & \ell_3 \\
  L''_3 & L''_2 & L''_1 \\
  1 & 1 & 2
  \end{array}
 \right\} \nonumber \\
&& \times 
\left[ \prod\limits^3_{n=1}
(-i)^{\ell_n}
\int_0^\infty {k_n^2 dk_n \over 2 \pi^2}
\mathcal{T}^{(Z_n)}_{X_n, \ell_n}(k_n)\right] \nonumber \\
&& \times 
\int_0^\infty A^2 dA j_{L_1} (k_1 A)  
\int_0^\infty B^2 dB j_{L'_1} (k_2 B)  
\int_0^\infty C^2 dC j_{L''_1} (k_3 C) \nonumber \\ 
&& \times \int_0^{k_D} k_1'^2 dk_1' P_B(k_1') 
j_{L_2} (k_1' A) j_{L'_3} (k_1' B) 
 \int_0^{k_D} k_2'^2 dk_2' P_B(k_2')
j_{L'_2} (k_2'B) j_{L''_3} (k_2'C) \nonumber \\
&& \times 
\int_0^{k_D} k_3'^2 dk_3'  P_B(k_3') 
j_{L''_2} (k_3'C) j_{L_3} (k_3'A) \nonumber \\
&& \times 
8 I_{L'_3 1 L}^{0 1 -1} I_{L_2 1 L}^{0 1 -1} I_{L''_3 1 L'}^{0 1 -1}
I_{L'_2 1 L'}^{0 1 -1} I_{L_3 1 L''}^{0 1 -1} I_{L''_2 1 L''}^{0 1 -1}
{\cal Q}_{L'_3, L_2, L} {\cal Q}_{L''_3, L'_2, L'} {\cal Q}_{L_3, L''_2,
L''}
\nonumber \\
&& \times 
2^{3-N_S} I_{L_1~\ell_1~2}^{0 |\lambda_1| -|\lambda_1|}
 I_{L_1'~\ell_2~2}^{0 |\lambda_2| -|\lambda_2|} I_{L_1''~\ell_3~2}^{0
 |\lambda_3| -|\lambda_3|} 
{\cal D}_{L_1, \ell_1, x_1}^{(e)} {\cal D}_{L_1', \ell_2, x_2}^{(e)} {\cal D}_{L_1'', \ell_3, x_3}^{(e)} 
~. \label{eq:CMB_bis_PMF_numerical}
\end{eqnarray}
Here, we define the filter functions as 
\begin{eqnarray}
\begin{split}
{\cal Q}_{L'_3, L_2, L} &\equiv (\delta_{L'_3, L+1} + \delta_{L'_3,
 |L-1|}) (\delta_{L_2, L+1} + \delta_{L_2, |L-1|}) + \delta_{L'_3, L}
 \delta_{L_2, L} \\
{\cal D}_{L_1, \ell_1, x_1}^{(e)} &\equiv (\delta_{L_1, \ell_1-2} +
 \delta_{L_1, \ell_1} + \delta_{L_1, \ell_1 + 2}) \delta_{x_1,0} 
+  (\delta_{L_1, \ell_1-1} + \delta_{L_1, \ell_1 + 1}) \delta_{x_1,1} ~,
\end{split}
\end{eqnarray}
which come from the selection rules of the Wigner symbols \cite{Shiraishi:2010kd}, and 
 \begin{eqnarray}
C_Z \equiv
\left\{
 \begin{array}{ll}
 - \frac{2}{\sqrt{3}} (8\pi)^{3/2} R_\gamma 
\ln \left(\tau_\nu / \tau_B\right)  & (Z = S) \\
 \frac{2}{3} (8\pi)^{3/2} & (Z = V) \\
  - 4 (8\pi)^{3/2} R_\gamma \ln \left( \tau_\nu / \tau_B \right) &
   (Z = T) 
 \end{array}
\right. ~.
\end{eqnarray}
This expression is used in the calculations of the exact bispectra presented in section~\ref{subsec:numerical_issues}.
 
\subsection{Pole approximation} \label{subsec:pole_app}

As we have mentioned in the previous subsection, it takes a lot of time to compute the CMB bispectrum based on the exact formula corresponding to the one-loop calculation. Therefore, in order to constrain on the PMFs through the estimation of the signal-to-noise ratio in a finite time, we find the reasonable expression of the CMB bispectrum by reducing the exact formula to the tree-level one. A similar approach can be seen in the calculation of the non-Gaussianity generated from scalar-type Gaussian-squared fields such as curvatons \cite{Lyth:2006gd}. 

Let us consider the structure of the bispectrum of the PMF anisotropic stresses (\ref{eq:bis_EMT}). This consists of three integrals over the dummy wave number vectors ${\bf k_1'}, {\bf k_2'}$, and ${\bf k_3'}$. If the magnetic spectrum is enough red as $n_B \sim -3$, these integrals are almost determined by the behavior of the integrand around at ``three poles'', namely, ${k_1'}, {k_2'}, {k_3'} \sim 0$. Picking up the effects around at these poles, we can express the bispectrum of the PMF anisotropic stresses with an asymptotic form as 
\begin{eqnarray}
\Braket{\Pi_{B ab}({\bf k_1}) \Pi_{B cd}({\bf k_2}) \Pi_{B ef}({\bf k_3})} 
&\sim& (- 4\pi \rho_{\gamma,0})^{-3} 
\frac{\alpha A_B}{n_B + 3} k_*^{n_B + 3} \frac{8\pi}{3}
 \delta\left(\sum_{n=1}^3 {\bf k_n}\right) \nonumber \\
&&\times 
\frac{1}{8}
\left[ P_B(k_1) P_B(k_2) \delta_{ad} P_{be}(\hat{\bf k_1}) P_{cf}(\hat{\bf k_2}) 
\right. \nonumber \\
&&\qquad \left.
+ P_B(k_2) P_B(k_3) P_{ad}(\hat{\bf k_2}) P_{be}(\hat{\bf k_3}) \delta_{cf} \right. \nonumber \\
&&\qquad \left. +  P_B(k_1) P_B(k_3) P_{ad}(\hat{\bf k_1}) \delta_{be} P_{cf}(\hat{\bf k_3}) 
\right. \nonumber \\
&&\qquad \left.
+ \{a \leftrightarrow b \ {\rm or} \ c \leftrightarrow d \ {\rm or} \ e \leftrightarrow f\} \right]~, 
\end{eqnarray}
where we have evaluated as 
\begin{eqnarray}
  \int d^3 {\bf k'} P_B(k') P_{ab}(\hat{\bf k'})
\sim 
\alpha \int_0^{k_*} k'^2 dk' P_B(k') \int d^2\hat{\bf k'} P_{ab}(\hat{\bf k'}) 
= \frac{\alpha A_B}{n_B + 3} k_*^{n_B + 3} \frac{8\pi}{3} \delta_{ab}
 ~. \label{eq:pole_app}
\end{eqnarray}
Here, we introduce $k_* = 10 {\rm Mpc}^{-1}$ and $\alpha$ as a cutoff scale of the integral and an unknown numerical factor which should be determined by the comparison with the exact bispectra described in the previous section, respectively\footnote{Equation~(\ref{eq:pole_app}) is valid for $n_B \neq -3$.}. 

Once we admit this validity, we formulate the four CMB bispectra generated from the scalar and tensor components of the PMF anisotropic stresses, namely the $SSS, SST, STT$, and $TTT$ spectra
\footnote{Of course, we also can apply the pole approximation to the CMB bispectra composed of the vector modes. However, we do not focus on these bispectra here since the contribution of the vector mode is subdominant for the scale where the current satellites are sensitive.}. 
The angular bispectrum of the primordial perturbations is approximately given by  
\begin{eqnarray}
\Braket{\prod_{n=1}^3 \xi_{\ell_n m_n}^{(\lambda_n)}(k_n)} 
&\sim& 
\left[\prod_{n=1}^3 \int d^2 \hat{\bf k_n} {}_{-\lambda_n}Y_{\ell_n m_n}^*(\hat{\bf k_n}) \right] 
\left[\frac{R_\gamma \ln(\tau_\nu / \tau_B )}{4\pi \rho_{\gamma,0}} \right]^3 \nonumber \\ 
&&\times \frac{\alpha A_B}{n_B + 3} k_*^{n_B + 3} 
 \frac{8\pi}{3} \delta\left(\sum_{n=1}^3 {\bf k_n}\right) \nonumber \\
&&\times 
\left[ P_B(k_1) P_B(k_2) \delta_{ad} P_{be}(\hat{\bf k_1}) P_{cf}(\hat{\bf k_2}) 
+ P_B(k_2) P_B(k_3) P_{ad}(\hat{\bf k_2}) P_{be}(\hat{\bf k_3}) \delta_{cf} \right. \nonumber \\
&&\quad \left. +  P_B(k_1) P_B(k_3) P_{ad}(\hat{\bf k_1}) \delta_{be} P_{cf}(\hat{\bf k_3}) \right]
 \nonumber \\
&&\times \left\{
 \begin{array}{ll}
 \frac{27}{8} O_{ab}^{(0)}(\hat{\bf k_1}) O_{cd}^{(0)}(\hat{\bf k_2})
 O_{ef}^{(0)}(\hat{\bf k_3}) & (\lambda_1 = \lambda_2 = \lambda_3 = 0) \\
- \frac{27}{4} O_{ab}^{(0)}(\hat{\bf k_1}) O_{cd}^{(0)}(\hat{\bf k_2}) e_{ef}^{(-\lambda_3)}(\hat{\bf k_3})  & (\lambda_1 = \lambda_2 = 0, ~ \lambda_3 = \pm 2) \\
 \frac{27}{2}  O_{ab}^{(0)}(\hat{\bf k_1}) e_{cd}^{(-\lambda_2)}(\hat{\bf k_2})
e_{ef}^{(-\lambda_3)}(\hat{\bf k_3}) & (\lambda_1 = 0, ~ \lambda_2, \lambda_3 = \pm 2) \\
 - 27 e_{ab}^{(-\lambda_1)}(\hat{\bf k_1}) e_{cd}^{(-\lambda_2)}(\hat{\bf k_2})
e_{ef}^{(-\lambda_3)}(\hat{\bf k_3}) & (\lambda_1, \lambda_2, \lambda_3 = \pm 2)
 \end{array}
\right. ~. \nonumber \\
\label{eq:ini_bis_optimal}
\end{eqnarray}
Using some relations in appendix~\ref{appen:polarization}, we reduce the contraction of the subscripts in the $SSS$ spectrum to 
\begin{eqnarray}
&& O_{ab}^{(0)}(\hat{\bf k_1}) O_{cd}^{(0)}(\hat{\bf k_2})
O_{ef}^{(0)}(\hat{\bf k_3})
\left[ P_B(k_1) P_B(k_2) \delta_{ad} P_{be}(\hat{\bf k_1}) P_{cf}(\hat{\bf k_2}) 
 \right. \nonumber \\
&&\qquad\qquad\qquad\qquad \left. + P_B(k_2) P_B(k_3) P_{ad}(\hat{\bf k_2}) P_{be}(\hat{\bf k_3}) \delta_{cf} +  P_B(k_1) P_B(k_3) P_{ad}(\hat{\bf k_1}) \delta_{be} P_{cf}(\hat{\bf k_3}) \right] \nonumber \\
&&\qquad\qquad =  \frac{8}{9} 
\left( \frac{4\pi}{3} \right)^3 
\sum_{L, L',L'' = 0,2} 
 \left\{
  \begin{array}{ccc}
  L & L' & L'' \\
  1 & 1 & 1
  \end{array}
 \right\} 
I_{L 1 1}^{0 1 -1} I_{L' 1 1}^{0 1 -1}  I_{L'' 1 1}^{0 1 -1}
\nonumber \\
&&\qquad\qquad\quad \times
\sum_{M M' M''} Y_{L M}^*(\hat{\bf k_1})
Y_{L' M'}^*(\hat{\bf k_2}) Y_{L'' M''}^*(\hat{\bf k_3}) 
\left(
  \begin{array}{ccc}
  L & L' & L'' \\
  M & M' & M''
  \end{array}
 \right) \nonumber \\
&&\qquad\qquad\quad \times
\left[ P_B(k_1) P_B(k_2) \delta_{L'', 2} + \delta_{L,2} P_B(k_2) P_B(k_3) 
+ P_B(k_1) \delta_{L', 2} P_B(k_3)\right]
~, 
\end{eqnarray}
that for the $SST$ spectrum to 
\begin{eqnarray}
&& O_{ab}^{(0)}(\hat{\bf k_1}) O_{cd}^{(0)}(\hat{\bf k_2}) e_{ef}^{(-\lambda_3)}(\hat{\bf k_3})
\left[ P_B(k_1) P_B(k_2) \delta_{ad} P_{be}(\hat{\bf k_1}) P_{cf}(\hat{\bf k_2}) 
 \right. \nonumber \\
&&\qquad\qquad\qquad\qquad \left. + P_B(k_2) P_B(k_3) P_{ad}(\hat{\bf k_2}) P_{be}(\hat{\bf k_3}) \delta_{cf} +  P_B(k_1) P_B(k_3) P_{ad}(\hat{\bf k_1}) \delta_{be} P_{cf}(\hat{\bf k_3}) \right] \nonumber \\
&&\qquad\qquad 
= - \frac{8\sqrt{3}}{9}
\left( \frac{4\pi}{3} \right)^3
\sum_{L,L',L'' = 0,2} 
I_{L 1 1}^{0 1 -1} I_{L' 1 1}^{0 1 -1} I_{L'' 1 1}^{0 1 -1} 
 \left\{
  \begin{array}{ccc}
  L & L' & L'' \\
  1 & 1 & 1
  \end{array}
 \right\} 
\nonumber \\
&&\qquad\qquad\quad \times 
\sum_{M M' M''} Y_{LM}^*(\hat{\bf k_1}) 
Y_{L' M'}^*(\hat{\bf k_2}) {}_{\lambda_3}Y_{L'' M''}^*(\hat{\bf k_3})
 \left(
  \begin{array}{ccc}
  L & L' & L'' \\
  M & M' & M''
  \end{array}
 \right) \nonumber \\
&&\qquad\qquad\quad \times \left[ P_B(k_1)P_B(k_2) + 3 \{ \delta_{L,2}P_B(k_2)  + P_B(k_1) \delta_{L',2} \} P_B(k_3) \right] \delta_{L'',2}
~, 
\end{eqnarray}
that for the $STT$ spectrum to 
\begin{eqnarray}
&& O_{ab}^{(0)}(\hat{\bf k_1}) e_{cd}^{(-\lambda_2)}(\hat{\bf k_2})
e_{ef}^{(-\lambda_3)}(\hat{\bf k_3})
\left[ P_B(k_1) P_B(k_2) \delta_{ad} P_{be}(\hat{\bf k_1}) P_{cf}(\hat{\bf k_2}) 
 \right. \nonumber \\
&&\qquad\qquad\qquad\qquad \left. + P_B(k_2) P_B(k_3) P_{ad}(\hat{\bf k_2}) P_{be}(\hat{\bf k_3}) \delta_{cf} +  P_B(k_1) P_B(k_3) P_{ad}(\hat{\bf k_1}) \delta_{be} P_{cf}(\hat{\bf k_3}) \right] \nonumber \\
&&\qquad\qquad =
8\left( \frac{4\pi}{3} \right)^3
\sum_{L,L',L''=0,2} I_{L 1 1}^{0 1 -1} I_{L' 1 1}^{0 1 -1} I_{L'' 1 1}^{0 1 -1} 
 \left\{
  \begin{array}{ccc}
  L & L' & L'' \\
  1 & 1 & 1
  \end{array}
 \right\} 
\nonumber \\
&&\qquad\qquad\quad \times 
\sum_{M M' M''} Y_{LM}^*(\hat{\bf k_1}) 
{}_{\lambda_2}Y_{L' M'}^*(\hat{\bf k_2}) {}_{\lambda_3}Y_{L'' M''}^*(\hat{\bf k_3})
 \left(
  \begin{array}{ccc}
  L & L' & L'' \\
  M & M' & M''
  \end{array}
 \right) \nonumber \\
&&\qquad\qquad\quad\times [P_B(k_1)\{ P_B(k_2) + P_B(k_3) \} + 3 \delta_{L,2} P_B(k_2) P_B(k_3) ] \delta_{L',2}\delta_{L'',2}
~, 
\end{eqnarray}
and that for the $TTT$ spectrum to 
\begin{eqnarray}
&&e_{ab}^{(-\lambda_1)}(\hat{\bf k_1}) e_{cd}^{(-\lambda_2)}(\hat{\bf k_2})
e_{ef}^{(-\lambda_3)}(\hat{\bf k_3})
\left[ P_B(k_1) P_B(k_2) \delta_{ad} P_{be}(\hat{\bf k_1}) P_{cf}(\hat{\bf k_2}) 
 \right. \nonumber \\
&&\qquad\qquad\qquad\qquad 
\left. + P_B(k_2) P_B(k_3) P_{ad}(\hat{\bf k_2}) P_{be}(\hat{\bf k_3}) \delta_{cf}
+  P_B(k_1) P_B(k_3) P_{ad}(\hat{\bf k_1}) \delta_{be} P_{cf}(\hat{\bf k_3}) \right] \nonumber \\
&&\qquad\qquad = 
- 24\sqrt{3} \left(\frac{4\pi}{3}\right)^3 \sum_{L,L',L'' = 0,2} I_{L 1 1}^{0 1 -1} I_{L' 1 1}^{0 1 -1} I_{L'' 1 1}^{0 1 -1} 
 \left\{
  \begin{array}{ccc}
   L & L' & L'' \\
  1 & 1 & 1
  \end{array}
 \right\} \nonumber \\
&&\qquad\qquad\quad\times 
\sum_{M, M', M''} {}_{\lambda_1} Y_{L M}^*(\hat{\bf k_1}) 
{}_{\lambda_2}Y_{L' M'}^*(\hat{\bf k_2}) {}_{\lambda_3}Y_{L'' M''}^*(\hat{\bf k_3})
\left(
  \begin{array}{ccc}
   L & L' & L'' \\
  M & M' & M''
  \end{array}
 \right) \nonumber \\
 &&\qquad\qquad\quad \times [P_B(k_1)P_B(k_2) + P_B(k_2)P_B(k_3) + P_B(k_1)P_B(k_3)]\delta_{L,2}\delta_{L',2}\delta_{L'',2}
~.
\end{eqnarray}
The delta function is also expanded with the spin spherical harmonics as equation~(\ref{eq:delta}). Then, the angular integrals of these spin spherical harmonics are performed as   
\begin{eqnarray}
\begin{split}
\int d^2 \hat{\bf k_1} {}_{-\lambda_1}Y_{\ell_1 m_1}^* Y_{L_1 M_1}^* {}_{\lambda_1}Y_{L M}^* &=
 I_{\ell_1 L_1 L}^{\lambda_1 0 -\lambda_1}
\left(
  \begin{array}{ccc}
  \ell_1 & L_1 & L \\
  m_1 & M_1 & M 
  \end{array}
 \right) ~, \\
\int d^2 \hat{\bf k_2} {}_{-\lambda_2}Y_{\ell_2 m_2}^* Y_{L_2 M_2}^* {}_{\lambda_2}Y_{L' M'}^* &= I_{\ell_2 L_2 L'}^{\lambda_2 0 -\lambda_2}
\left(
  \begin{array}{ccc}
  \ell_2 & L_2 & L' \\
  m_2 & M_2 & M' 
  \end{array}
 \right) ~, \\
\int d^2 \hat{\bf k_3} {}_{-\lambda_3}Y_{\ell_3 m_3}^* Y_{L_3 M_3}^* {}_{\lambda_3}Y_{L'' M''}^* &= I_{\ell_3 L_3 L''}^{\lambda_3 0 -\lambda_3}
\left(
  \begin{array}{ccc}
  \ell_3 & L_3 & L'' \\
  m_3 & M_3 & M'' 
  \end{array}
 \right) ~,
\end{split}
\end{eqnarray}
and resulting Wigner-$3j$ symbols are summed up as  
\begin{eqnarray}
&& \sum_{\substack{M_1 M_2 M_3 \\ M M' M''}}
\left(
  \begin{array}{ccc}
  L_1 &  L_2 & L_3 \\
   M_1 & M_2 & M_3
  \end{array}
 \right)
\left(
  \begin{array}{ccc}
  L &  L' & L'' \\
   M & M' & M''
  \end{array}
 \right) 
\left(
  \begin{array}{ccc}
  \ell_1 &  L_1 & L \\
   m_1 & M_1 & M
  \end{array}
 \right)
\left(
  \begin{array}{ccc}
  \ell_2 &  L_2 & L' \\
   m_2 & M_2 & M'
  \end{array}
 \right)
\left(
  \begin{array}{ccc}
  \ell_3 &  L_3 & L'' \\
   m_3 & M_3 & M''
  \end{array}
 \right) \nonumber \\
&& \qquad\qquad\qquad = 
\left(
  \begin{array}{ccc}
  \ell_1 & \ell_2 & \ell_3 \\
   m_1 & m_2 & m_3
  \end{array}
 \right)
\left\{
  \begin{array}{ccc}
  \ell_1 & \ell_2 & \ell_3 \\
   L_1 & L_2 & L_3 \\
   L & L' & L'' \\
  \end{array}
 \right\}~. 
\end{eqnarray} 
Thus, the resultant initial bispectrum (\ref{eq:ini_bis_optimal}) are rewritten as 
\begin{eqnarray}
\Braket{\prod_{n=1}^3 \xi_{\ell_n m_n}^{(\lambda_n)}(k_n)} 
&\sim& 
\left(
  \begin{array}{ccc}
  \ell_1 & \ell_2 & \ell_3 \\
   m_1 & m_2 & m_3
  \end{array}
 \right) 
\left[\frac{R_\gamma \ln(\tau_\nu / \tau_B )}{4\pi \rho_{\gamma,0}} \right]^3 
\frac{\alpha A_B}{n_B + 3} k_*^{n_B + 3} \frac{8\pi}{3} \nonumber \\
&&\times
8 \int_0^\infty y^2 dy 
\left[ \prod_{n=1}^3 \sum_{L_n} 
 (-1)^{L_n/2} j_{L_n}(k_n y) \right]  
I_{L_1 L_2 L_3}^{0 \ 0 \ 0} 
\nonumber \\
&&\times  
\left( \frac{4\pi}{3} \right)^3
\sum_{L, L',L'' = 0,2} 
 \left\{
  \begin{array}{ccc}
  L & L' & L'' \\
  1 & 1 & 1
  \end{array}
 \right\} 
I_{L 1 1}^{0 1 -1} I_{L' 1 1}^{0 1 -1} I_{L'' 1 1}^{0 1 -1}  \nonumber \\
&&\times 
I_{\ell_1 L_1 L}^{\lambda_1 0 -\lambda_1}
I_{\ell_2 L_2 L'}^{\lambda_2 0 -\lambda_2} I_{\ell_3 L_3 L''}^{\lambda_3 0 -\lambda_3}
\left\{
  \begin{array}{ccc}
  \ell_1 & \ell_2 & \ell_3 \\
   L_1 & L_2 & L_3 \\
   L & L' & L'' \\
  \end{array}
 \right\}
 \nonumber \\
&&\times 
 \begin{cases}
\parbox{6.5cm}{%
 \flushleft %
$ 
3 [ P_B(k_1) P_B(k_2) \delta_{L'', 2}$ \\ 
\quad $ + \delta_{L,2} P_B(k_2) P_B(k_3) $ \\ 
\quad $+ P_B(k_1) \delta_{L', 2} P_B(k_3) ]  
$ }  
& \parbox{4cm}{%
 \flushleft %
$ (\lambda_1 = \lambda_2 = \lambda_3 = 0) 
$ } 
\\
\parbox{6.5cm}{%
 \flushleft %
$ 6\sqrt{3} 
[  P_B(k_1) P_B(k_2) $ \\ 
\qquad $+ 3 \{ \delta_{L,2} P_B(k_2)+ P_B(k_1) \delta_{L', 2} \}$ \\  
\qquad\quad $\times P_B(k_3)
] \delta_{L'',2}
$} 
& \parbox{4cm}{%
 \flushleft %
$ 
(\lambda_1 = \lambda_2 = 0, ~ \lambda_3 = \pm 2) 
$ }
\\
 \parbox{6.5cm}{%
 \flushleft %
 $  108  [ P_B(k_1) \{ P_B(k_2) + P_B(k_3) \} $ \\
\qquad$ + 3 \delta_{L,2} P_B(k_2)P_B(k_3) ] \delta_{L',2} \delta_{L'',2}
$ } 
& 
\parbox{4cm}{%
 \flushleft %
$ 
(\lambda_1 = 0, ~ \lambda_2, \lambda_3 = \pm 2) 
$ } 
\\
 \parbox{6.5cm}{%
 \flushleft %
$ 
648\sqrt{3} [ P_B(k_1)P_B(k_2) + P_B(k_2)P_B(k_3) $ \\ 
\qquad\quad$+ P_B(k_1)P_B(k_3) ] 
\delta_{L,2} \delta_{L',2} \delta_{L'',2}
$} 
& 
\parbox{4cm}{%
 \flushleft %
$ 
(\lambda_1, \lambda_2, \lambda_3 = \pm 2) 
$} 
 \end{cases}
. \nonumber \\
\end{eqnarray}
Substituting the above expression into equation (\ref{eq:CMB_bis_PMF_general}) and using the result of the summation over $\lambda_1, \lambda_2$ and $\lambda_3$ (\ref{eq:summation_helicities}), 
the resultant approximate CMB bispectrum composed of the scalar and tensor modes is derived:
\begin{eqnarray}
B_{X_1 X_2 X_3, \ell_1 \ell_2 \ell_3}^{{\rm app}~(Z_1 Z_2 Z_3)}(\alpha)
&=& 
\left[\frac{R_\gamma \ln(\tau_\nu / \tau_B )}{4\pi \rho_{\gamma,0}} \right]^3 
\frac{\alpha A_B}{n_B + 3} k_*^{n_B + 3} \frac{8\pi}{3}  
\sum_{L_1 L_2 L_3} (-1)^{\frac{L_1 + L_2 + L_3}{2}} I_{L_1 L_2 L_3}^{0 \ 0 \ 0}
\nonumber \\
&&\times
\left( \frac{4\pi}{3} \right)^3
\sum_{L, L', L'' = 0,2} 
 \left\{
  \begin{array}{ccc}
  L & L' & L'' \\
  1 & 1 & 1
  \end{array}
 \right\} 
I_{L 1 1}^{0 1 -1} I_{L' 1 1}^{0 1 -1} I_{L'' 1 1}^{0 1 -1} 
\left\{
  \begin{array}{ccc}
  \ell_1 & \ell_2 & \ell_3 \\
   L_1 & L_2 & L_3 \\
   L & L' & L'' \\
  \end{array}
 \right\}
\nonumber \\
&&\times 
2^{3 - N_S}I_{\ell_1 L_1 L}^{|\lambda_1| 0 -|\lambda_1|} I_{\ell_2 L_2 L'}^{|\lambda_2| 0 -|\lambda_2|} I_{\ell_3 L_3 L''}^{|\lambda_3| 0 -|\lambda_3|} 
{\cal D}_{L_1, \ell_1, x_1}^{(e)} {\cal D}_{L_2, \ell_2, x_2}^{(e)} {\cal D}_{L_3, \ell_3, x_3}^{(e)} 
\nonumber \\
&&\times 
8 \int_0^\infty y^2 dy 
\left[ \prod_{n=1}^3 
 (-i)^{\ell_n} \int_0^\infty \frac{k_n^2 dk_n}{2 \pi^2} 
{\cal T}_{X_n, \ell_n}^{(Z_n)}(k_n) j_{L_n}(k_n y) \right]  \nonumber \\
&&\times 
 \begin{cases}
\parbox{6.5cm}{%
 \flushleft %
$ 
3 [ P_B(k_1) P_B(k_2) \delta_{L'', 2}$ \\ 
\quad $ 
+ \delta_{L,2} P_B(k_2) P_B(k_3) $ \\ 
\quad $+ P_B(k_1) \delta_{L', 2} P_B(k_3) ]  
$ }  
& \parbox{4cm}{%
 \flushleft %
$ (Z_1 = Z_2 = Z_3 = S) 
$ } 
\\
\parbox{6.5cm}{%
 \flushleft %
$ 6\sqrt{3} 
[  P_B(k_1) P_B(k_2) $ \\ 
\qquad $+ 3 \{ \delta_{L,2} P_B(k_2)+ P_B(k_1) \delta_{L', 2} \}$ \\  
\qquad\quad $\times P_B(k_3)
] \delta_{L'',2}
$} 
& \parbox{4cm}{%
 \flushleft %
$ 
(Z_1 = Z_2 = S, ~ Z_3 = T) 
$ }
\\
 \parbox{6.5cm}{%
 \flushleft %
 $  108  [ P_B(k_1) \{ P_B(k_2) + P_B(k_3) \} $ \\
\qquad$ + 3 \delta_{L,2} P_B(k_2)P_B(k_3) ] \delta_{L',2} \delta_{L'',2}
$ } 
& 
\parbox{4cm}{%
 \flushleft %
$ 
(Z_1 = S, ~ Z_2 = Z_3 = T) 
$ } 
\\
 \parbox{6.5cm}{%
 \flushleft %
$ 
648\sqrt{3} [ P_B(k_1)P_B(k_2) + P_B(k_2)P_B(k_3) $ \\ 
\qquad\quad$+ P_B(k_1)P_B(k_3) ] 
\delta_{L,2} \delta_{L',2} \delta_{L'',2}
$} 
& 
\parbox{4cm}{%
 \flushleft %
$ 
(Z_1 = Z_2 = Z_3 = T) 
$} 
 \end{cases}
. \label{eq:cmb_bis_PMF_pole_app} \nonumber \\
\end{eqnarray}
Comparing the above expression with the exact form of the CMB bispectrum [(\ref{eq:CMB_bis_PMF_complete}) or (\ref{eq:CMB_bis_PMF_numerical})], we can see that the numbers of the time integrals and dummy multipoles decrease. In addition, note that the dependence on $k_1, k_2$ and $k_3$ is identical to that in the CMB bispectrum arising from the local-type non-Gaussianity of curvature perturbations which is given by
\begin{eqnarray}
B_{X_1 X_2 X_3, \ell_1 \ell_2 \ell_3}^{\rm local}
&=& I_{\ell_1 \ell_2 \ell_3}^{0~0~0}
\int_0^\infty y^2 dy 
\left[ \prod_{n=1}^3 \frac{2}{\pi} \int_0^\infty k_n^2 dk_n 
{\cal T}_{X_n,\ell_n}^{(S)}(k_n) j_{\ell_n}(k_n y) \right] 
\nonumber \\
&&\times 
\frac{6}{5} f_{\rm NL}^{\rm local} 
\left[ P_\zeta(k_1) P_\zeta(k_2) + {\rm 2 \ perms.} \right] 
\label{eq:cmb_bis_local}
\end{eqnarray}
with $f_{\rm NL}^{\rm local}$ and $P_\zeta$ being the so-called nonlinearity parameter of the local-type non-Gaussianity and the power spectrum of curvature perturbations, respectively. These imply that by the pole approximation, the one-loop calculation (the left panel of figure~\ref{fig:diagram}) changes to the tree-level one (the right one of figure~\ref{fig:diagram}). Thus, we can reduce the time cost involved with calculating the bispectrum. 

\subsection{Numerical issues}\label{subsec:numerical_issues}

\begin{figure}[t]
  \begin{tabular}{cc}
    \begin{minipage}{0.5\hsize}
  \begin{center}
    \includegraphics[width=7.5cm,clip]{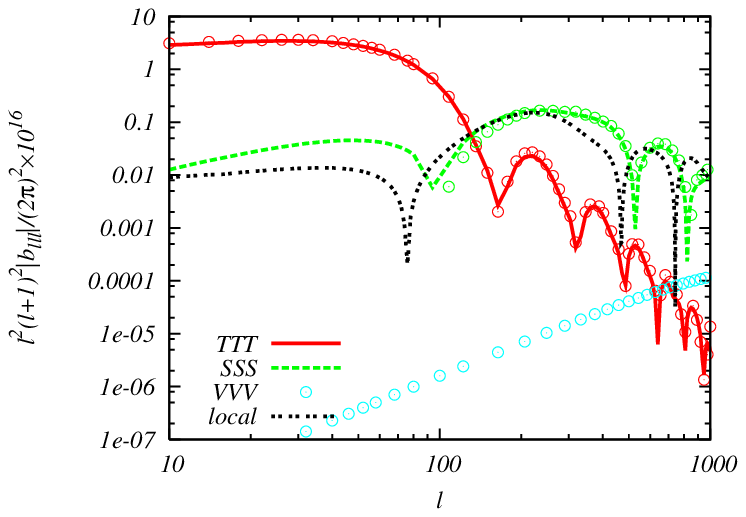}
  \end{center}
\end{minipage}
\begin{minipage}{0.5\hsize}
  \begin{center}
    \includegraphics[width=7.5cm,clip]{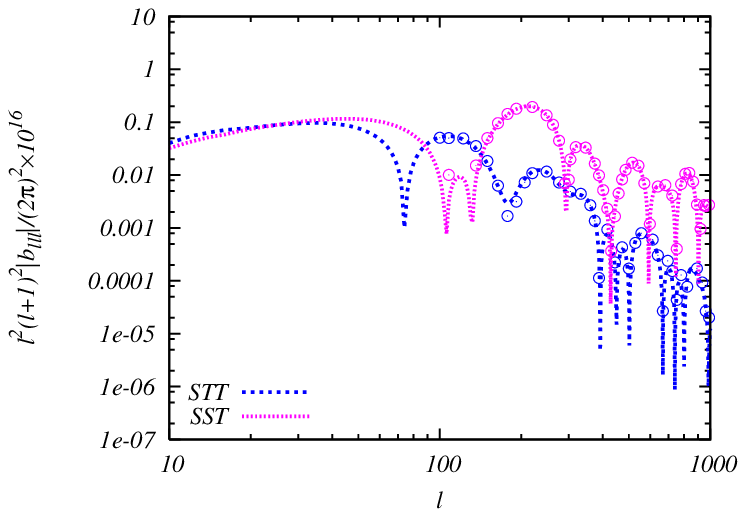}
  \end{center}
\end{minipage}
\end{tabular}
  \caption{Absolute values of the CMB bispectra of the intensity fluctuations generated from the magnetic $SSS, SST, STT, TTT$ and $VVV$ modes, and the local-type non-Gaussianity of curvature perturbations (\ref{eq:cmb_bis_local}) with $f_{\rm NL}^{\rm local} = 5$ for $\ell_1 = \ell_2 = \ell_3 \equiv \ell$. The circle points and lines of the magnetic $SSS, SST, STT, TTT$ and $VVV$ modes correspond to the exact bispectra (\ref{eq:CMB_bis_PMF_numerical}) and the approximate ones (\ref{eq:cmb_bis_PMF_pole_app}), respectively. The PMF parameters are fixed to $B_{1 \rm Mpc} = 4.7 {\rm nG}, n_B = -2.9$ and $\tau_\nu / \tau_B = 10^{17}$, and other parameters are identical to the mean values derived from the WMAP 7-yr data \cite{Komatsu:2010fb}} \label{fig:PMF_bis}
\end{figure}

Here, we present the numerical results of the exact and approximate bispectra for $n_B = -2.9$ based on equations~(\ref{eq:CMB_bis_PMF_numerical}) and (\ref{eq:cmb_bis_PMF_pole_app}). To compute the actual bispectra, we modified the Boltzmann Code for Anisotropies in the Microwave Background \cite{Lewis:1999bs, Lewis:2004ef} and use the Common Mathematical Library SLATEC \cite{slatec}. 

In figure~\ref{fig:PMF_bis}, we plot the reduced bispectra of the intensity modes, which is given by
\begin{eqnarray}
b_{I I I, \ell_1 \ell_2 \ell_3}^{(Z_1 Z_2 Z_3)} \equiv (I_{\ell_1 \ell_2 \ell_3}^{0~0~0})^{-1} B_{I I I, \ell_1 \ell_2 \ell_3}^{(Z_1 Z_2 Z_3)}~,
\end{eqnarray}
on the basis of the exact (circle points) and approximate (lines) formulae of the magnetic case for $(Z_1, Z_2, Z_3)$ = $(S, S, S)$, $(S, S, T)$, $(S, T, T)$, $(T, T, T)$, $(V, V, V)$, and the non-magnetic formula (\ref{eq:cmb_bis_local}). 
From this figure, we can see that the approximate $SSS, SST, STT$ and $TTT$ spectra (\ref{eq:cmb_bis_PMF_pole_app}) are in good agreement with the exact ones (\ref{eq:CMB_bis_PMF_numerical}). Quantitatively, using the correlation function for each mode $b \cdot b' \equiv \sum_{\ell} b_{III, \ell \ell \ell}^{(Z_1 Z_2 Z_3)} b_{III, \ell \ell \ell}'^{(Z_1 Z_2 Z_3)}$, the correlation coefficients between the exact ($b^{\rm ex}$) and approximate ($b^{\rm app}$) bispectra are evaluated as 
\begin{eqnarray}
 \frac{b^{\rm ex} \cdot b^{\rm app}}
{ \sqrt{ \left( b^{\rm ex} \cdot b^{\rm ex} \right)
\left( b^{\rm app} \cdot b^{\rm app} \right) } }
= \begin{cases}
 0.994 & (Z_1 = Z_2 = Z_3 = S) \\
 0.999 & (Z_1 = Z_2 = S, ~ Z_3 = T) \\ 
 0.996 & (Z_1 = S, ~ Z_2 = Z_3 = T) \\
 0.9997 & (Z_1 = Z_2 = Z_3 = T)
\end{cases}~.
\end{eqnarray}
These become almost one, therefore in the following discussion, we believe that equation~(\ref{eq:cmb_bis_PMF_pole_app}) reconstructs the shapes of the exact bispectra. 

The remaining factors, $\alpha$'s, which are introduced in equation~(\ref{eq:pole_app}), are determined by an asymptotic relation as
\begin{eqnarray}
\alpha = \frac{b^{\rm ex}}{b^{\rm app}(\alpha = 1)}
\approx  \frac{b^{\rm ex} \cdot b^{\rm app}(\alpha = 1)}{b^{\rm app}(\alpha = 1) \cdot b^{\rm app}(\alpha = 1)}
=  
\begin{cases}
 0.3350 & (Z_1 = Z_2 = Z_3 = S) \\
 0.3473 & (Z_1 = Z_2 = S, ~ Z_3 = T) \\ 
 0.3212 & (Z_1 = S,~ Z_2 = Z_3 = T) \\
 0.2991 & (Z_1 = Z_2 = Z_3 = T)
\end{cases} .
\end{eqnarray}
These values are used in figure~\ref{fig:PMF_bis} and in the estimation of the signal-to-noise ratio described in the next section. 

From here, let us discuss the behaviors of these bispectra. On large scales ($\ell \lesssim 100$), the tensor mode is the most significant signal in the intensity anisotropy due to the boost of the Integrated Sachs-Wolfe effect \cite{Pritchard:2004qp, Shiraishi:2011dh}, and highly damped at $\ell \gtrsim 100$. On the other hand, the scalar mode stands out at $\ell \gtrsim 100$ owing to the acoustic oscillation of the baryon-photon fluid. Although the damping by the viscosity of photons is effective for $\ell \gtrsim 1000$, the scalar mode keeps on dominating the tensor one \cite{Shaw:2009nf}. The vector mode is monotonically increasing for $\ell \lesssim 2000$ due to the vorticity of photons enhanced by the Lorentz force \cite{Durrer:1998ya, Mack:2001gc}. Regardless of it, this acts as a subdominant signal for $\ell \lesssim 1000$ \cite{Shaw:2009nf}. 
These features can be seen in the CMB bispectra. That is, from figure~\ref{fig:PMF_bis}, it is observed that the more tensor modes the bispectrum comes from, the larger scale it shows up on. In this figure, the contribution of the vector mode seems to be negligible. 
In addition, we also notice that the magnetic $SSS$ bispectrum and the bispectrum from the local-type non-Gaussianity of curvature perturbations have the similar shapes with each other despite the different dependence on multipoles. This backs up the fact that the non-Gaussianity of the PMF anisotropic stresses is classified into the local-type configuration. 

\section{Signal-to-noise ratio}\label{sec:SN}

In this section, using an approximate formula for the CMB bispectrum generated from the scalar and tensor perturbations of the PMF anisotropic stresses (\ref{eq:cmb_bis_PMF_pole_app}), we compute the signal-to-noise ratio and estimate the optimal observational bound on the PMF strength when the PMF spectrum is nearly scale invariant as $n_B = -2.9$. 

According to refs.~\cite{Luo:1993xx, Komatsu:2001rj, Bartolo:2004if}, the signal-to-noise ratio with respect to the CMB bispectra is formulated as 
\begin{eqnarray} 
\left(\frac{S}{N}\right)^2 =  \sum_{2 \leq \ell_1 \leq \ell_2 \leq \ell_3 \leq \ell_{\rm max}} 
\frac{\left[\sum_{Z_1 Z_2 Z_3} B_{III, \ell_1 \ell_2 \ell_3}^{(Z_1 Z_2 Z_3)}\right]^2}
{\Delta_{\ell_1 \ell_2 \ell_3} \prod_{n=1}^3 
\left( C_{\ell_n}^{\rm fid} + N_{\ell_n} \right) }~, \label{eq:SN_bis}
\end{eqnarray}
where $\Delta_{\ell_1 \ell_2 \ell_3} = 1~({\rm for \ \ell_1 \neq \ell_2 \neq \ell_3}), ~ 6~ ({\rm for \ \ell_1 = \ell_2 = \ell_3}), ~ 2 ~ ({\rm otherwise})$ denotes a numerical factor caused by the cosmic variance, $C_{\ell}^{\rm fid}$ is the fiducial CMB power spectrum which is consistent with the current observational data \cite{Komatsu:2010fb}, and $N_\ell$ is the noise spectrum given by \cite{Luo:1993xx, Knox:1995dq}
\begin{eqnarray}
N_\ell &=&  
\left( \theta_{\rm FWHM} \frac{\Delta T}{2.726 \rm K} \right)^2 e^{\ell(\ell+1) \theta_{\rm FWHM}^2 / (8 \ln 2)}~,
\end{eqnarray}
with $\theta_{\rm FWHM}$ and $\Delta T$ being the full width at half-maximum of the Gaussian beam and the instrumental noise per pixel in terms of temperature, respectively. 

\begin{figure}[t]
  \centering \includegraphics[width=10cm,clip]{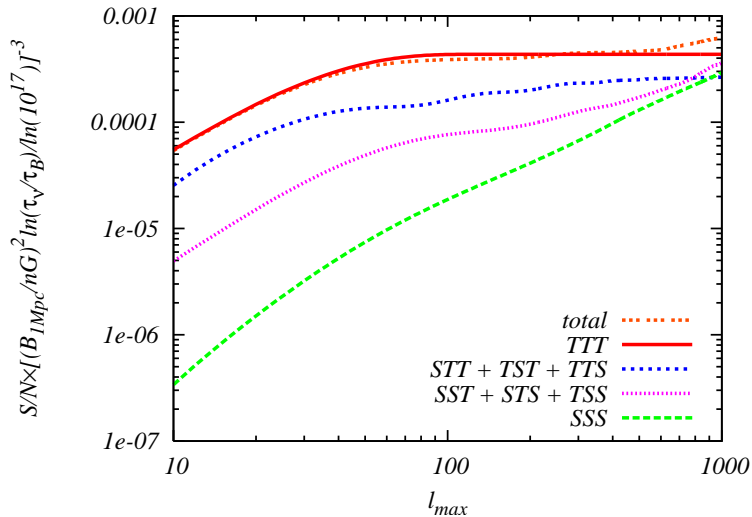}
  \caption{Noise-free signal-to-noise ratios. The ``total'' line denotes $S/N$ obtained from the total spectrum of the $SSS, SST, STS, TSS, STT, TST, TTS$ and $TTT$ spectra, and the others correspond to $S/N$'s coming from each mode. Here, we fix the PMF parameters as $n_B = -2.9$. The other parameters are identical to the mean values obtained from the WMAP-7yr data \cite{Komatsu:2010fb}.}
  \label{fig:SN_III_S+T_B4.7_n-2.9_tntb1e17.eps}
\end{figure}

Firstly, as an ideal limit, we describe the signal-to-noise ratios derived from the $SSS$ (green line), $SST+STS+TSS$ (magenta one), $STT+TST+TTS$ (blue one) and $TTT$ (red one) spectra and their total spectrum (orange one) when $N_{\ell} = 0$ in figure~\ref{fig:SN_III_S+T_B4.7_n-2.9_tntb1e17.eps}. From these, we can analyze the behavior of the total signal-to-noise ratio: at first, owing to the boost of the $TTT$ spectrum, the total signal-to-noise ratio rapidly increases for $\ell_{\rm max} \lesssim 100$. Below that angular scale, the contribution of the $TTT$ spectrum is completely saturated for $\ell_{\rm max} \gtrsim 100$, and the cross-correlated bispectra between the scalar and tensor modes lead to the gradual growth. If $\ell_{\rm max}$ reaches $700$, the contribution of the $SSS$ spectrum becomes effective and the total signal-to-noise ratio starts to increase again. 
One may wonder about a fact that $S/N$ from the total CMB spectrum is smaller than that from the $TTT$ spectrum at $\ell_{\rm max} \sim 100$. This is due to the compensation of the total CMB bispectrum induced by the other-mode spectra. 

\begin{figure}[t]
  \centering \includegraphics[width=10cm,clip]{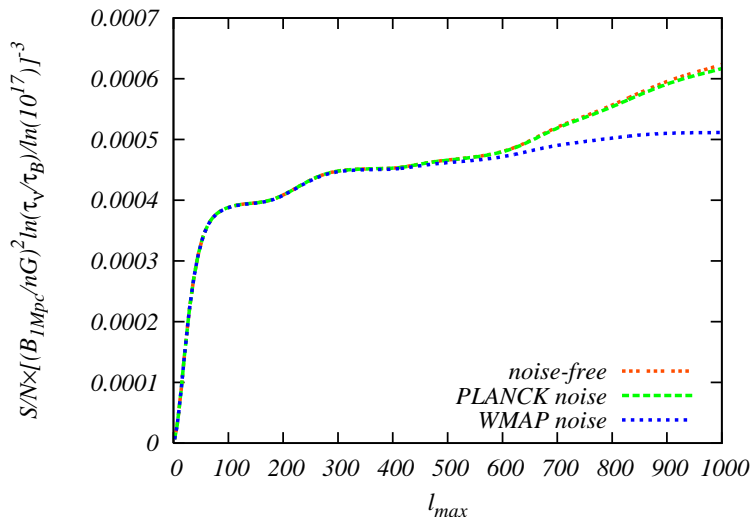}
  \caption{Signal-to-noise ratios with zero, PLANCK- and WMAP-level noises obtained from the auto- and cross-correlated bispectra between the scalar and tensor modes. The instrumental noises and resolutions of the WMAP and PLANCK experiments are given by $(\theta_{\rm FWHM}{[\rm arcmin]}, \Delta T{[\rm \mu K]}) = (13,18.3), (5.0, 13.1)$, respectively \cite{Smith:2006ud, Xia:2007gz, Hu:2001fb}. Here, we fix the PMF parameters to the values presented in figure~\ref{fig:SN_III_S+T_B4.7_n-2.9_tntb1e17.eps} and the other parameters are identical to the mean values obtained from the WMAP-7yr data \cite{Komatsu:2010fb}.}
  \label{fig:SN_III_S+T_B4.7_n-2.9_tntb1e17_noise.eps}
\end{figure}

Figure~\ref{fig:SN_III_S+T_B4.7_n-2.9_tntb1e17_noise.eps} shows the signal-to-noise ratios if we take into account the finite instrumental noises and angular resolutions in the WMAP and PLANCK experiments, respectively as $(\theta_{\rm FWHM}{[\rm arcmin]}, \Delta T{[\rm \mu K]}) = (13,18.3), (5.0, 13.1)$ \cite{Smith:2006ud, Xia:2007gz, Hu:2001fb}. 
We can see that the signal-to-noise ratios with the WMAP and PLANCK noises start to peak out from $\ell \sim 500$ and $\sim 1000$, respectively. 
From this figure, we shall find the observational limits on the PMF magnitude for $n_B = -2.9$. At $\ell_{\rm max} = 1000$, the signal-to-noise ratios with the WMAP and PLANCK noises, which are normalized by $B_{1 \rm Mpc} = 1 {\rm nG}$ and $\tau_\nu / \tau_B = 10^{17}$, are $5.13 \times 10^{-4}$ and $6.16 \times 10^{-4}$, respectively. Then, equating these values with $S/N \propto (B_{1 \rm Mpc})^6 [\ln (\tau_\nu / \tau_B)]^3$, we can see that $S/N < 2$ and $\tau_\nu / \tau_B = 10^{17} - 10^6$ lead to the upper bounds on the PMF strength at $95\%$ confidence level: $B_{1 \rm Mpc} < 4.0 - 6.7 {\rm nG} ~ {\rm (WMAP \ limit)}, 3.8 - 6.5 {\rm nG} ~ {\rm (PLANCK \ limit)}$. 
One may be disappointed that these bounds are somewhat relaxed compared with the previous ones estimated from only the $SSS$ or $TTT$ spectrum \cite{Trivedi:2010gi, Shiraishi:2011dh}. However, these were based on a rough estimation and may involve some uncertainties, hence, our new results are more reliable. 

\section{Summary and discussion}

In this paper, we newly studied how the PMF strength is constrained from the auto- and cross-correlated bispectra between the CMB intensity fluctuations of the scalar and tensor passive modes in the case where the PMF spectrum closes to the scale invariance as $n_B = -2.9$. To do this, at first, we derived the most general formula for the CMB bispectrum generated from the bispectrum of the PMF anisotropic stresses including all considerable correlations with no approximation in accordance with the computation approach of ref.~\cite{Shiraishi:2011fi}. However, due to the dependence of the bispectrum of the PMF anisotropic stresses on the six-point correlation of the Gaussian PMFs, this formula involves a huge computational effort corresponding to the one-loop diagram. To avoid this difficulty, by evaluating the behaviors of the integrands with respect to the dummy wave number vectors around the poles, we approximately expressed the bispectrum of the PMF anisotropic stresses with the four-point correlation of the Gaussian variables and reduced the formula to the tree-level bispectrum. 
In this attempt, we also confirmed that the non-Gaussianity originated from the PMF anisotropic stresses has the local-type configuration. 
Through the comparison with the exact spectra, we checked that this approximate formula appropriately produces the exact shape. 

Using the approximate formula, we estimated the signal-to-noise ratio. For the noise-free case, we analyzed the behavior of this and checked the consistency of our calculation. Considering the instrumental noise and angular resolutions of the WMAP experiment, we obtained a bound on the PMF strength at 95\% confidence level: $B_{1 \rm Mpc} < 4.0 - 6.7 {\rm nG}$, where the upper and lower values correspond to the cases of $\tau_\nu / \tau_B = 10^{17}, 10^6$. In the PLANCK experiment, this value becomes slightly tigher as $B_{1 \rm Mpc} < 3.8 - 6.5 {\rm nG}$. 
One can notice that these limits are weaker than the recent constraint and PLANCK forecast by the CMB power spectra reported in ref.~\cite{Paoletti:2010rx}. The analysis in ref.~\cite{Paoletti:2010rx} also includes information about the polarizations and contributions of the vector mode, which we do not consider here. Constraining from the polarization bispectrum induced by PMFs will be an interesting subject and one of our future works.

In the previous attempts about probing the PMF with the CMB bispectrum \cite{Seshadri:2009sy, Caprini:2009vk, Trivedi:2010gi}, there exist many uncertainties in the evaluation of the CMB bispectra. In our recent works \cite{Shiraishi:2010yk, Shiraishi:2011fi, Shiraishi:2011dh}, regardless of the precise treatments of the CMB bispectra, the bounds have been roughly estimated. In this sense, we note that this is the first paper in which the properties of the PMF are extracted from the CMB bispectrum in the appropriate way. 
This study will be extended to the higher-order correlations such as the trispectrum \cite{Trivedi:2011vt}. If this realizes, we will acquire deeper comprehensions of the nature of the PMF. 

\acknowledgments
This work was supported in part by a Grant-in-Aid for JSPS Research under Grant No.~22-7477 (M. S.), JSPS Grant-in-Aid for Scientific Research under Grant No.~22340056 (S. Y.), No.~22012004 (K. I.), Grant-in-Aid for Scientific Research on Priority Areas No.~467 ``Probing the Dark Energy through an Extremely Wide and Deep Survey with Subaru Telescope'', and Grant-in-Aid for Nagoya University Global COE Program ``Quest for Fundamental Principles in the Universe:
 from Particles to the Solar System and the Cosmos'', from the Ministry
 of Education, Culture, Sports, Science and Technology of Japan. 
We also acknowledge the Kobayashi-Maskawa Institute for the Origin of Particles and the Universe, Nagoya University, for providing computing resources useful in conducting the research reported in this paper.  

\appendix
\section{Projection tensors}\label{appen:polarization}

Here, we briefly summarize the useful properties of the projection tensors of the scalar, vector and tensor modes on the basis of ref.~\cite{Shiraishi:2010kd}. 

At first, we define an arbitrary unit vector, polarization vector and tensor, which are expanded by the spin spherical harmonics, as 
\begin{eqnarray}
\begin{split}
\hat{k}_a &= \sum_m \alpha_a^{m} Y_{1 m}(\hat{\bf k}) ~, \\
\epsilon_a^{(\pm 1)} (\hat{\bf k}) 
&= \mp \sum_m \alpha_a^m {}_{\pm 1} Y_{1 m} (\hat{\bf k})~, \\
e_{ab}^{(\pm 2)} (\hat{\bf k}) 
&=\sqrt{2} \epsilon_a^{(\pm 1)} (\hat{\bf k}) \epsilon_b^{(\pm 1)} (\hat{\bf k}) ~,
\end{split}
\end{eqnarray}
with
\begin{eqnarray}
\alpha_a^m \equiv \sqrt{\frac{2 \pi}{3}}
 \left(
  \begin{array}{ccc}
   -m (\delta_{m,1} + \delta_{m,-1}) \\
   i~ (\delta_{m,1} + \delta_{m,-1}) \\
   \sqrt{2} \delta_{m,0}
  \end{array}
\right)~.
\end{eqnarray}
The scalar product of $\alpha_a^m$, which is required for the tensor contraction, is easily calculated as
\begin{eqnarray}
\alpha_a^m \alpha_a^{m'} = \frac{4 \pi}{3} (-1)^m \delta_{m,-m'}~, \ \
\alpha_a^m \alpha_a^{m' *} = \frac{4 \pi}{3} \delta_{m,m'}~.
\end{eqnarray}
  
These polarization vector and tensor obey the divergenceless and transverse-traceless conditions and some properties in the Fourier space as 
\begin{eqnarray}
\begin{split}
\hat{k}^a \epsilon_a^{(\pm 1)}(\hat{\bf k}) &= 0~, \\
\epsilon^{(\pm 1) *}_a (\hat{\bf k}) &= \epsilon^{(\mp 1)}_a (\hat{\bf k})
 = \epsilon^{(\pm 1)}_a (-\hat{\bf k})~, \\
\epsilon^{(\lambda)}_a (\hat{\bf k}) \epsilon^{(\lambda')}_a (\hat{\bf k}) 
&= \delta_{\lambda, -\lambda'} \ \ \ ({\rm for} \ \lambda, \lambda' = \pm 1) ~, \\
e_{aa}^{(\pm 2)}(\hat{\bf k}) &= \hat{k}_a e_{ab}^{(\pm 2)}(\hat{\bf k}) = 0~, \\
e_{ab}^{(\pm 2) *}(\hat{\bf k}) &= e_{ab}^{(\mp 2)}(\hat{\bf k}) = e_{ab}^{(\pm
2)}(- \hat{\bf k})~, \\
e_{ab}^{(\lambda)}(\hat{\bf k}) e_{ab}^{(\lambda')}(\hat{\bf k}) &= 2
\delta_{\lambda, -\lambda'} \ \ \ ({\rm for} \ \lambda, \lambda' = \pm 2)~. \label{eq:pol_tens_relation}
\end{split}
\end{eqnarray}
Using these vectors and tensor, an arbitrary physical tensor, such as the metric or the PMF anisotropic stress, is decomposed into the two scalar ($\chi_{\rm iso}, \chi^{(0)}$), two vector ($\chi^{(\pm 1)}$) and two tensor ($\chi^{(\pm 2)}$) components:
\begin{eqnarray}
\chi_{ab}({\bf k}) &=& - \frac{1}{3} \chi_{\rm iso}({\bf k}) \delta_{ab} + \chi^{(0)}({\bf k}) O^{(0)}_{ab}(\hat{\bf k}) \nonumber \\
&& + \sum_{\lambda = \pm 1} \chi^{(\lambda)}({\bf k}) 
\left[ \hat{k}_a \epsilon^{(\lambda)}_b(\hat{\bf k}) + \hat{k}_b \epsilon^{(\lambda)}_a(\hat{\bf k}) \right]
+ \sum_{\lambda = \pm 2} \chi^{(\lambda)}({\bf k}) e^{(\lambda)}_{ab}(\hat{\bf k})~,
\end{eqnarray}
where $O^{(0)}_{ab}(\hat{\bf k}) \equiv - \hat{k}_a \hat{k}_b + \delta_{ab} / 3$.
Considering equation~(\ref{eq:pol_tens_relation}), we have the inverse formulae as 
\begin{eqnarray}
\begin{split}
\chi^{(0)}({\bf k}) &= \frac{3}{2}O^{(0)}_{ab}(\hat{\bf k}) \chi_{ab}({\bf k}) ~, \\
\chi^{(\pm 1)}({\bf k}) &= \frac{1}{2} 
\left[ \hat{k}_a \epsilon_b^{(\mp 1)}(\hat{\bf k}) 
+ \hat{k}_b \epsilon_a^{(\mp 1)}(\hat{\bf k}) \right] \chi_{ab}({\bf k})~, \\ 
\chi^{(\pm 2)}({\bf k}) &= \frac{1}{2} e_{ab}^{(\mp 2)}(\hat{\bf k}) \chi_{ab}({\bf k})~. \label{eq:tensor_inverse}
\end{split}
\end{eqnarray}

From these, we find several relations of the projection tensors as
\begin{eqnarray}
\begin{split}
O_{ab}^{(0)}(\hat{\bf k}) &= - 2 I_{2 1 1}^{0 1 -1} \sum_{M m_a m_b} 
Y_{2 M}^*(\hat{\bf k}) \alpha_a^{m_a} \alpha_b^{m_b} 
\left(
  \begin{array}{ccc}
  2 & 1 & 1 \\
  M & m_a & m_b 
  \end{array}
 \right) ~, \\
P_{ab}(\hat{\bf k}) &\equiv \delta_{ab} - \hat{k}_a \hat{k}_b 
= \sum_{\sigma = \pm 1} \epsilon_a^{(\sigma)}(\hat{\bf k}) \epsilon_b^{(-\sigma)}(\hat{\bf k}) \\
&= -2 \sum_{L=0,2} I_{L 1 1}^{0 1 -1} \sum_{M m_a m_b} 
Y^*_{L M}(\hat{\bf k}) \alpha_a^{m_a} \alpha_b^{m_b} 
\left(
  \begin{array}{ccc}
  L & 1 & 1 \\
  M & m_a & m_b
  \end{array}
 \right) ~, \\
\hat{k}_a {\epsilon}_b^{(\pm 1)}(\hat{\bf k}) 
+ \hat{k}_b {\epsilon}_a^{(\pm 1)}(\hat{\bf k}) 
&= \pm 2\sqrt{3} I_{2 1 1}^{0 1 -1} \sum_{M m_a m_b} 
{}_{\mp 1} Y^*_{2 M}(\hat{\bf k}) \alpha_a^{m_a} \alpha_b^{m_b} 
\left(
  \begin{array}{ccc}
  2 & 1 & 1 \\
  M & m_a & m_b
  \end{array}
 \right)~, \\
e_{ab}^{(\pm 2)}(\hat{\bf k}) &= 2 \sqrt{3} I_{2 1 1}^{0 1 -1} 
\sum_{M m_a m_b} {}_{\mp 2}Y_{2 M}^*(\hat{\bf k})
\alpha_a^{m_a} \alpha_b^{m_b}
 \left(
  \begin{array}{ccc}
  2 & 1 & 1 \\
  M & m_a & m_b
  \end{array}
 \right) ~, \\
O_{ab}^{(0)}(\hat{\bf k}) P_{bc}(\hat{\bf k}) 
&= \frac{1}{3} P_{ac}(\hat{\bf k}) 
~, \\
e^{(\pm 2)}_{ab}(\hat{\bf k}) P_{bc}(\hat{\bf k}) 
&= e_{ac}^{(\pm 2)}(\hat{\bf k}) ~. \label{eq:projection_tensors}
\end{split}
\end{eqnarray}
These representations are useful for the tensor contractions under the exact and pole approximation as shown in sections~\ref{subsec:exact} and \ref{subsec:pole_app}. 

\bibliography{paper}
\end{document}